\documentclass[twocolumn,times]{aastex62}

\defcitealias{warren2018}{Paper I}

\shorttitle{The Magnetic Properties of Heating Events}
\shortauthors{Ugarte-Urra, Crump, Warren \& Wiegelmann}

\begin{document}


\title{The Magnetic Properties of Heating Events on High-Temperature Active Region Loops}

\author[0000-0001-5503-0491]{Ignacio Ugarte-Urra}
\affil{Space Science Division, Naval Research Laboratory, Washington, DC 20375}

\author{Nicholas A. Crump}
\affil{Space Science Division, Naval Research Laboratory, Washington, DC 20375}

\author[0000-0001-6102-6851]{Harry P. Warren}
\affil{Space Science Division, Naval Research Laboratory, Washington, DC 20375}

\author{Thomas Wiegelmann}
\affil{Max-Planck-Institut f{\"u}r Sonnensystemforschung, Justus-von-Liebig-Weg 3, 37077 G{\"o}ttingen, Germany}


\begin{abstract}
  Understanding the relationship between the magnetic field and coronal heating is
one of the central problems of solar physics. However, studies of the magnetic
properties of impulsively heated loops have been rare. We present results from a
study of 34 evolving coronal loops observed in the \ion{Fe}{18} line component of
AIA/{\it SDO} 94 \AA\ filter images from three active regions with different magnetic
conditions.  We show that the peak intensity per unit cross-section of the loops
depends on their individual magnetic and geometric properties. The intensity scales
proportionally to the average field strength along the loop ($B_{avg}$) and
inversely with the loop length ($L$) for a combined dependence of $(B_{avg}/L)^{0.52\pm0.13}$.
These loop properties are inferred from magnetic
extrapolations of the photospheric HMI//{\it SDO} line-of-sight and vector magnetic
field in three approximations: potential and two Non Linear Force-Free
(NLFF) methods. Through hydrodynamic modeling (EBTEL model) we show that this behavior
is compatible with impulsively heated loops with a volumetric heating rate that
scales as $\epsilon_H\sim B_{avg}^{0.3\pm0.2}/L^{0.2\pm^{0.2}_{0.1}}$.

\end{abstract}



\section{introduction}
Understanding the relationship between the magnetic field and coronal heating is
one of the central problems of solar physics. It is well established that in the
regions on the Sun where the surface magnetic fields are stronger, the plasma in
the atmosphere reaches higher temperatures and emits larger amounts of ultraviolet
and X-ray radiation.
Global quantities such as the luminosity or the total radiance
scale up with the total unsigned magnetic flux of the regions through power-law
relationships \citep[e.g.][]{schrijver1987,fisher1998,benevolenskaya2002,fludra2002,pevtsov2003,fludra2008}.

The atmosphere, frozen-in to the magnetic field, responds to the long term changes
that the surface flows impose on the magnetic elements. Over time, major reorganization
of the fields result in follow-up changes in the morphology of the coronal
emission and the radiance \citep[e.g.][]{vanDriel2015,vanDriel2003,demoulin2004,ugarte-urra2015}.

At the time scale of minutes, the magnetic field distributions evolve slowly. Subject
to flows of $\sim$1$\rm\,km\,s^{-1}$ \citep[e.g.][]{berger1996}, a 5\arcsec\ flux
element experiences a modest displacement of about its own size within an hour.
In that interval, coronal emission in an active region can evolve rather quickly
with as many as five heating events as observed in spectral lines formed around
the 3--5 MK temperature \citep{ugarte-urra2014}. The atmospheric response at the
chromospheric footpoints of the loops is even more rapid \citep{testa2014}. The
challenge for coronal heating is, therefore, understanding the coupling between
the forcing that convection motions impose to the magnetic field and the response
of the atmosphere.

As surface flows carry the energy, their presence provides clues about where this energy
can be deposited. For example, in the umbra of sunspots flows are inhibited \citep[e.g.][]{borrero2011},
therefore it is not surprising to see most coronal loops rooted outside them.
\citet{tiwari2017} have shown that the occasional loops observed with one footpoint
rooted in umbra have the opposite footpoint anchored in plage or penumbra. Field
line tangling followed by reconnection \citep[see review][]{klimchuk2015}, and
wave dissipation \citep[see review][]{arregui2015} have been the traditional
heating scenarios considered for the transfer of energy. The observational constraints,
however, are yet insufficient to rule out any of these models.

In this study, we investigate the coupling between magnetic fields and radiance.
Our goal is to provide insight on how the magnitude of the heating scales up with the
properties of the local magnetic field once the heating mechanism is already in
place. There are two key properties of the field that are important to this
coupling: magnetic field strength ($B$) and loop geometry ($L$). This is discussed
in detail in \citet{mandrini2000}, where following the theoretical implications
of SXT/{\it Yohkoh} loop diagnostics discussed by \citet{porter1995}, they compile
a list of potential heating mechanisms and how each depends on quantities
such as magnetic field strength, loop length, velocity and density. While, in principle,
this implies that observational constraints on the scaling can help us discriminate
between models, the effectiveness of this method as a final discriminator is still
not guaranteed \citep[see discussion in][]{mandrini2000,klimchuk2006}

The studies that have looked into these constraints, nevertheless, have demonstrated that
such a  parameterization of heating can be very successful in reproducing several
aspects of coronal emission.
Forward modeling of active regions and the full Sun as an ensemble of hydrodynamic
loops using a volumetric heating rate that scales as a function of the average
magnetic field strength in the loop and its length ($\epsilon_H\propto B_{avg}^{\alpha} / L^{\beta}$)
has been successful in reproducing relationships for global quantities. Several of these
studies find $\alpha=1$ and $\beta=1$ to be a good prescription to reproduce
the emission of hot ($\gtrsim3$ MK) loops at the core \citep{warren2006,warren2007,lundquist2008b,ugarte-urra2017},
but they have problems replicating the emission of cooler 1-2 MK emission. \citet{winebarger2008}
showed that attempting to match the cool moss emission can lead to a different
scaling ($B_{avg}^{0.3} / L$). Furthermore, modeling the full Sun, \citet{schrijver2004}
found a power-law dependence ($B_0 / L^2$) on footpoint field strength $B_0$
\citep[see also][]{dudik2011}, which \citet{warren2006} argued to be consistent
as they obtained $B_{avg}\sim B_0/L$.

In our study, we revisit this problem by taking advantage on the progress made in
recent years in our ability to first observe loops at high spatial resolution, high
temporal resolution and temperature discrimination, and then model their magnetic
topology with improved magnetic models of the atmosphere. The objectives are two-fold:
obtain novel constraints on how observational properties like the radiance at high
temperatures for specific stages in the evolution of an individual loop scale with
$B$ and $L$; and relate those results to the demands imposed on the volumetric
heating rate.

In Section~\ref{sect:methods} we describe the dataset and methods used to extract
the magnetic and radiative properties of 34 coronal loops from three active regions.
Section~\ref{sect:results} presents the results of comparing the two properties for
all loops. The implications of those results for coronal heating are discussed in
Section~\ref{sect:discussion}. We find that, at the scale of individual loops, the
intensities from the \ion{Fe}{18} 93.93 \AA\ indeed depend on the individual properties
of the loops and scale proportionally with the average
field strength of the loop and inversely with its length. We provide new
constraints for models of heating in the corona.

\section{Methods}
\label{sect:methods}
This paper is a follow up to \citet{warren2018}, from now on \citetalias{warren2018},
an investigation that compared loops traces in coronal images with field lines computed
from various magnetic field extrapolation methods. \citetalias{warren2018} used a list of 15 active
regions, first compiled in a study of high temperature emission in active region cores
\citep{warren2012}. The current investigation extends the analysis of the magnetic
topology to find a link to the plasma properties of selected loops in a subset of the
active regions. Where not explicitly stated, the observations and methods
are the same as those described in \citetalias{warren2018}.

\subsection{Dataset}
Our dataset consists of 34 loops from a subset of three active regions (NOAA 11158,
11190, 11339) from the active region list in \citetalias{warren2018}. The full list
includes regions that span one order of magnitude in total unsigned magnetic flux.
These three active regions were selected because they cover most of the magnetic
flux range ($4.2\times10^{21}$Mx, $1.8\times10^{22}$Mx, $2.6\times10^{22}$Mx, respectively);
and a visual inspection of the coronal images shows well defined loop
structures that are suitable for the identification, manipulation and quantitative
analysis described in Section~\ref{sect:methods}.

For each active region we downloaded cutouts from the {\it Solar Dynamics Observatory} \citep{pesnell2012}
Joint Science Operations Center\footnote{http://jsoc.stanford.edu/} for a one-hour interval.
The downloads included data from the Atmospheric Imaging Assembly \citep[AIA,][]{lemen2012}
at 12\,s cadence in several EUV channels, plus line-of-sight and vector magnetograms
from the Helioseismic and Magnetic Imager \citep[HMI,][]{scherrer2012} at 45\,s
and 720\,s cadence, respectively.

\begin{figure*}[t!]
  \centerline{%
  \includegraphics[clip=true, width=0.33\textwidth]{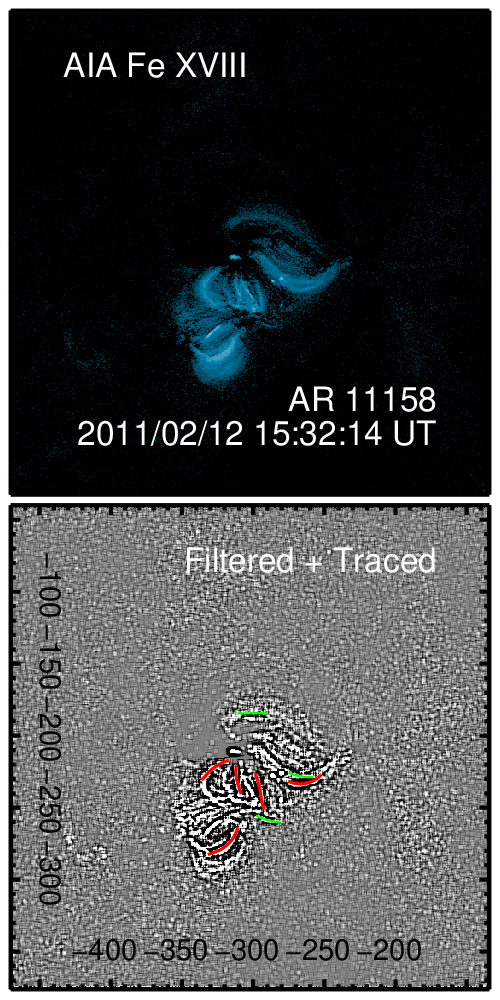}
  \includegraphics[clip=true, width=0.33\textwidth]{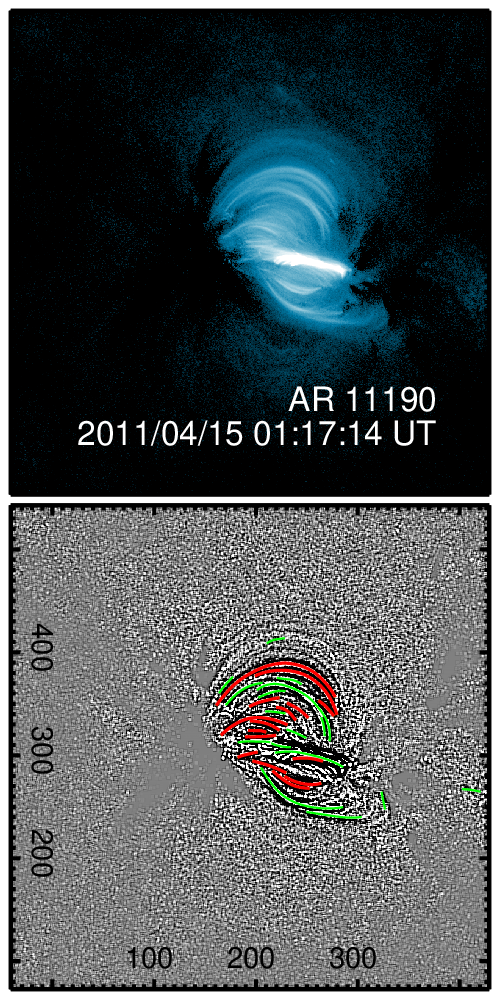}
  \includegraphics[clip=true, width=0.33\textwidth]{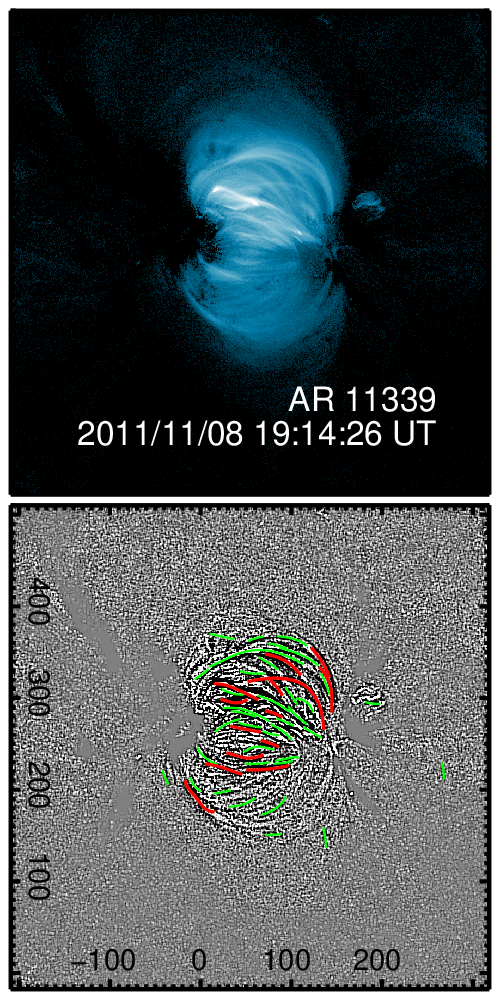}
  }
  \caption{Coronal loops identified using the automated loop tracing algorithm for all three active regions.
  The top row shows the AIA \ion{Fe}{18} eight-image average that served as input. All regions are shown at
  the same logarithmic intensity scaling. The bottom row shows the
  resulting filtered image with the automatically identified traced segments on top. The red lines represent
  the subset of the segments selected for the full topological and radiative analysis. Solar coordinates are
  provided in arcseconds from disk center.
  }
\label{fig:loopselect}
\end{figure*}

Unlike \citetalias{warren2018} that looks at the loops in several of the AIA bandpasses,
in this study, we only considered what we call AIA \ion{Fe}{18} images. These are images
from the 94\,\AA\ channel that have been processed to isolate and retain the emission
from the \ion{Fe}{18} 93.93 \AA\ spectral line, while removing the contribution from cooler lines
\citep[see all contributions in][]{odwyer2010}. The empirical correction
\citep[equation 1 in][]{ugarte-urra2014} was devised by \citet{warren2012}. We
chose \ion{Fe}{18} because we consider it an optimal choice for plasma diagnostics
in active region loops with current instrumentation. \ion{Fe}{18} images can be
obtained frequently and at high resolution with AIA. The spectral line forms within
the temperature range 3 –- 7 MK where active region cores peak in their emission
measure \citep{warren2012}. Furthermore, we have shown that the intensity peak in
this line takes place before the loop reaches the equilibrium point, when radiation
starts to dominate the cooling in an impulsively heated loop \citep{ugarte-urra2017}.
Observing before the start of the radiative cooling phase is important in
our ability to diagnose the heating properties \citep{winebarger2004}.

\subsection{Loop identification}
Loops in the AIA \ion{Fe}{18} images were identified using the Oriented Coronal CUrved
Loop Tracing (OCCULT-2) algorithm \citep{aschwanden2010,aschwanden2013}. The algorithm uses
a low-pass filter to eliminate noise and a high-pass filter to enhance the fine structure.
Combined through subtraction they operate as an unsharp masking filter that reveals ridges
that can be traced. Starting from the highest contrast locations, loops are identified,
and then erased from the images to allow the procedure to be repeated, and loops accumulated until
reaching the threshold level for detection.

Following \citetalias{warren2018}, and to improve signal-to-noise, we applied this method
to an eight-image average (96 s interval) at the middle time of the
observing sequence for all three active regions. Our choice of time is different
than \citetalias{warren2018}, where the start time was selected. This change is motivated
by our need to investigate the loop evolution before and after the time of identification.
We identified 8, 31, and 51 loops in NOAA 11158, 11190 and 11339 respectively. Note that
the number of identifications correlates with the total \ion{Fe}{18} emission in the regions \citep{warren2012}.
The traced loop segments are shown in Figure~\ref{fig:loopselect}.
This number is larger than the number of loops used in our full analysis because we
introduce further constraints in the selection based on the analysis of the topology and the
intensity evolution. We provide further details in sections~\ref{sect:extrap} -- ~\ref{sect:intens}.

\begin{figure*}[t!]
  \centerline{%
  \includegraphics[clip=true, width=0.33\textwidth]{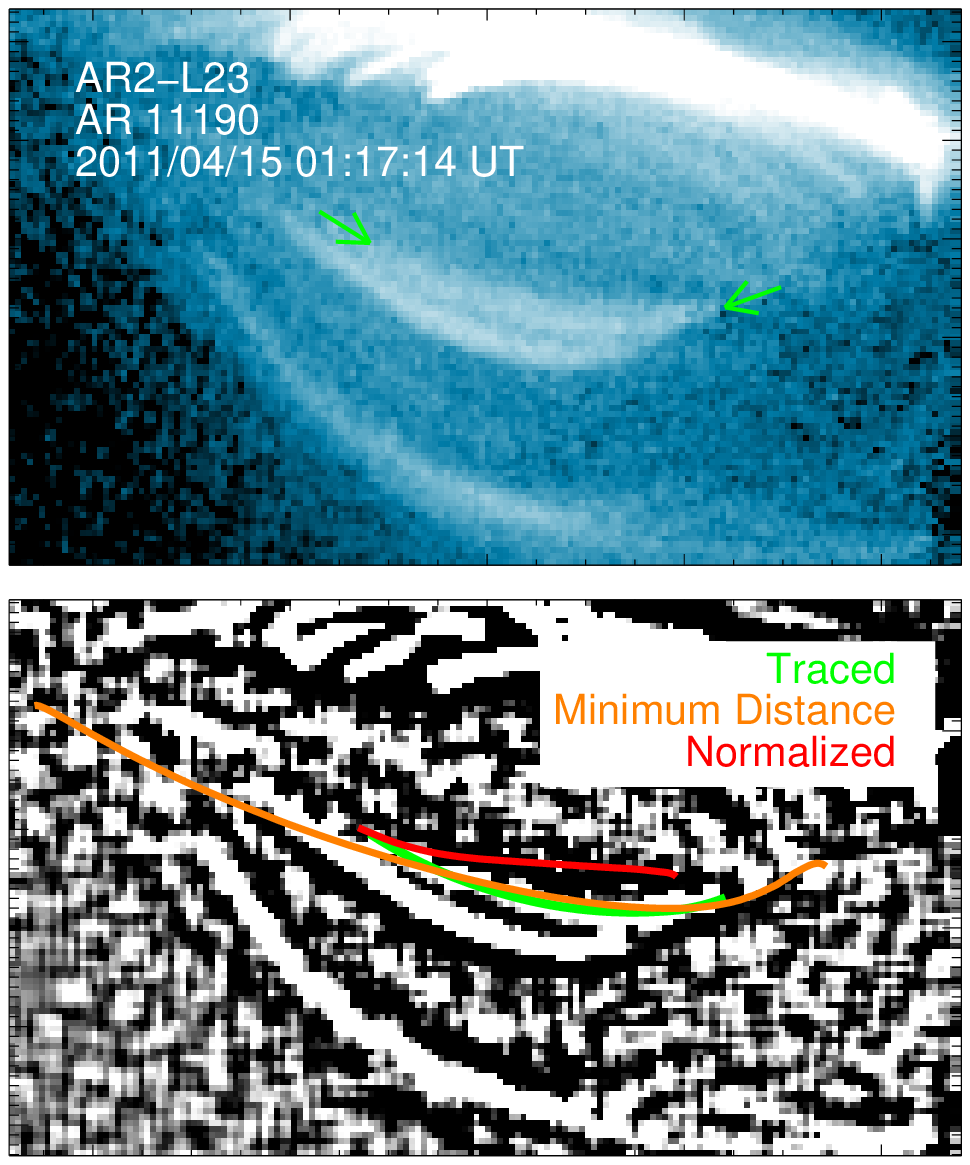}
  \includegraphics[clip=true, width=0.33\textwidth]{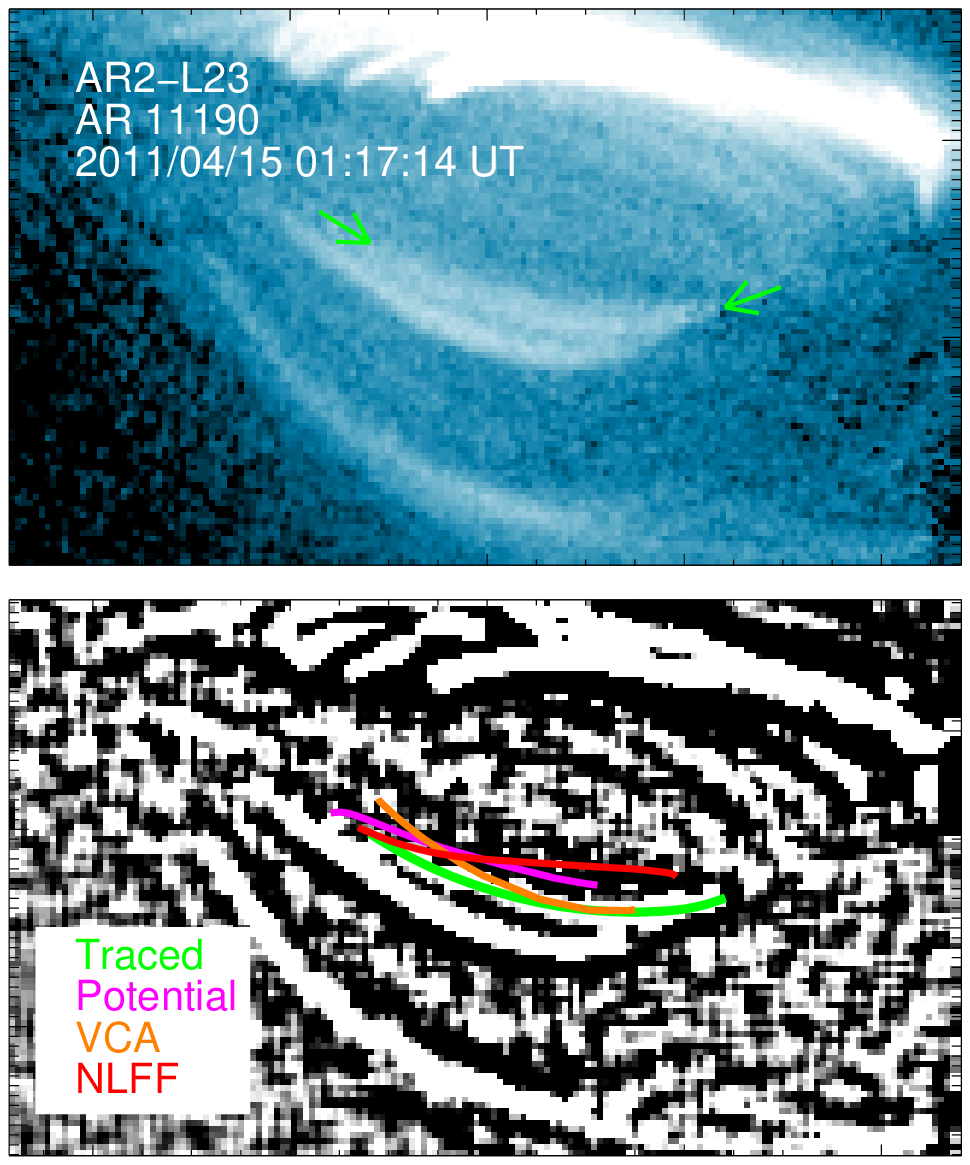}
  \includegraphics[clip=true, width=0.33\textwidth]{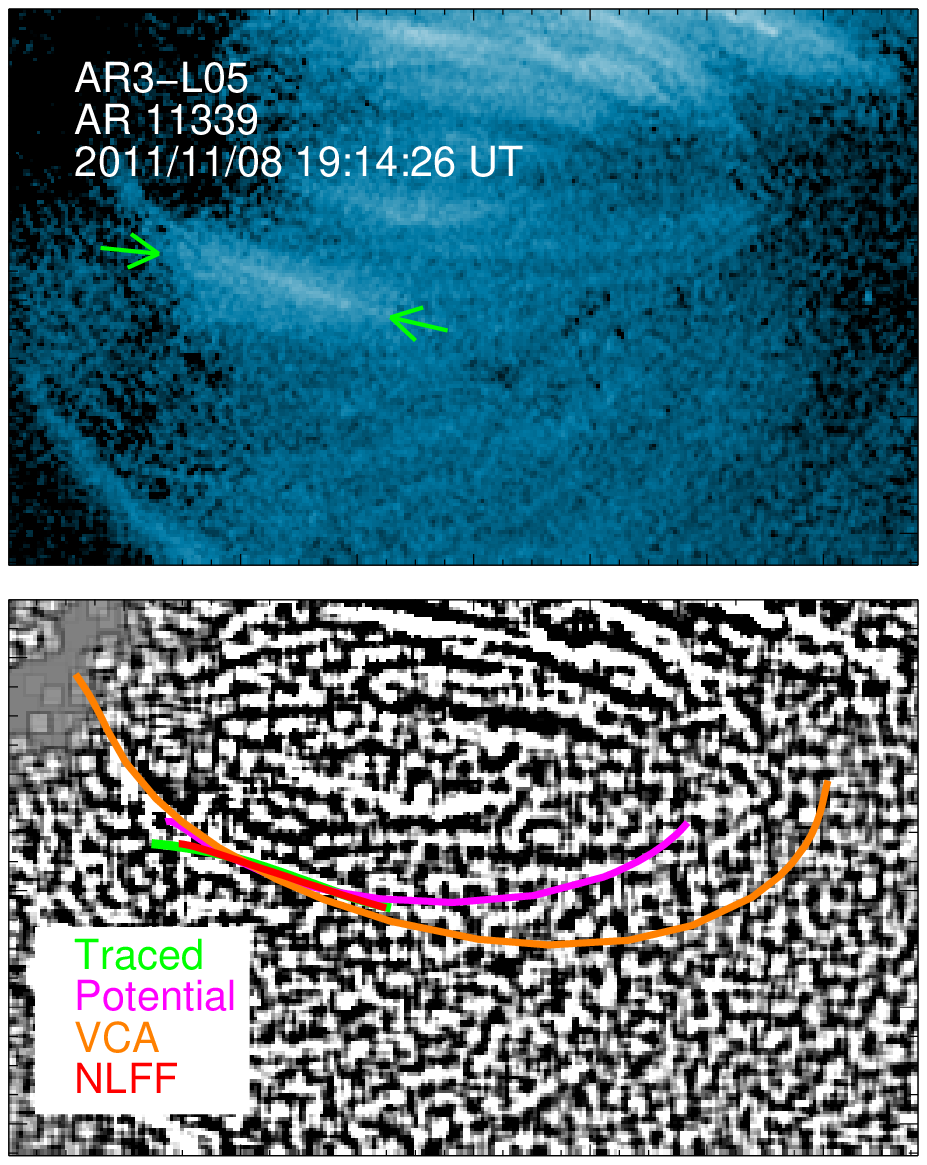}
  }
  \caption{The left panel shows a loop where the best field line match for
  a traced loop is better achieved by using a metric where distances along the loop
  and field line are first normalized by distance to one footpoint. The middle panel shows
  differences in the best field line match for that same loop and three extrapolation methods
  (potential, VCA and NLFF). These differences can sometimes be significant, as the right panel
  shows for a different loop in a different active region. In those cases inspecting the
  temporal evolution of the loop helps understanding what model works best for that traced loop.
  The green arrows indicate the location of the traced loop.
  }
\label{fig:loopmethods}
\end{figure*}

\subsection{Magnetic field extrapolations}
\label{sect:extrap}
The magnetic properties of the loops were determined from magnetic models of the
active regions. We considered three magnetic field extrapolation techniques: a simple
potential field extrapolation,  the Non Linear Force-Free (NLFF) field extrapolation method of
\citet{wiegelmann2012}, and the Vertical-Current Approximation VCA-NLFF method
described in \citet{aschwanden2016}. The three methods are described in some detail in
\citetalias{warren2018}, here we summarize the basic assumptions.

The potential approximation assumes a current-free volume and solves the field equations
($\nabla\times\mathbf{B} = 0$, $\nabla\cdot\mathbf{B} = 0$) using the corrected
line-of-sight component of the observed magnetic field as a lower boundary. Solutions are
obtained  by means of Fourier transforms as in our previous studies of the topology of
active regions in connection to coronal heating \citep{warren2006,ugarte-urra2017} and
the initiation of coronal mass ejections \citep{ugarte-urra2007}. We use HMI line-of-sight
magnetograms as the boundary layer.

Force-free magnetic models do allow currents in the volume, but neglect all non-magnetic forces
\citep[e.g.][]{wiegelmann2012b}. The field equations are solved assuming a vanishing Lorentz
force $\mathbf{J}\times\mathbf{B} = 0$, that leaves us with $\nabla\times\mathbf{B} = {4\pi \over c} \mathbf{J} = \alpha \mathbf{B}$,
and $\nabla\cdot\mathbf{B} = 0$. This approximations is called Non-Linear Force-Free (NLFF),
and the twist parameter $\alpha$ changes throughout the volume as $\mathbf{B}\cdot\nabla\alpha=0$.
We obtain solutions using the code developed by \citet{wiegelmann2012}. As a constraint we use
HMI photospheric  vector magnetic fields that are preprocessed to create a
boundary condition that is consistent with the force-free field assumption.

The VCA-NLFF code uses the line-of-sight magnetogram as the boundary and  assumes that
the magnetic field can be decomposed into a superposition
of magnetic charges below the surface with a non-potential field prescribed by a helical twist.
The code optimizes the non-potential parameters of the model by comparing the field
lines to the traced loops obtained from a call to the OCCULT-2 loop identification
algorithm. As in \citetalias{warren2018}, and only for this module, we use data from all
the following AIA channels: 94, 131, 171, 193, 211, 335\,\AA.

For each active region we use the extrapolations calculated for \citetalias{warren2018} at the
start of the one hour observing sequence. The output of
the three methods is the magnetic field decomposed in Cartesian coordinates. Field lines were
calculated for seeds from all magnetogram pixels with a magnetic flux density above 25 G. The
field line integrator uses a fourth-order Runge-Kutta
method with adaptive step size, written in C.

\begin{deluxetable*}{ccccccccccccccDDc}
\tabletypesize{\small}
\tablecaption{Loops dataset.
The magnetic properties listed here are derived from the selected best-fit field
line of the three extrapolation models. The \ion{Fe}{18} properties are derived
from the observed loops identified in the AIA images. The loops have a code identifier that
links them to one of the three active regions in the sample, and the list of loop segments
identified for that region. The magnetic properties for the best-fit solution
of all methods are underlined. The properties are in bold font for the field line
selected from visual inspection of the best-fits. \label{tab:loops}}
\tablewidth{0pt}
\tablehead{
\multicolumn{1}{c}{} &
\multicolumn{1}{c}{} &
\multicolumn{1}{c}{} &
\multicolumn{1}{c}{} &
\multicolumn{1}{c}{} &
\multicolumn{7}{c}{Magnetic Properties} &
\multicolumn{1}{c}{} &
\multicolumn{6}{c}{\ion{Fe}{18} Properties} \\  \cline{6-12}  \cline{14-19}
\multicolumn{1}{c}{}    &
\multicolumn{1}{c}{} &
\multicolumn{2}{c}{Loop}  &
\multicolumn{1}{c}{} &
\multicolumn{4}{c}{$B_{avg}$ [G]} &
\multicolumn{3}{c}{Length [\arcsec]} &
\multicolumn{1}{c}{} &
\multicolumn{1}{c}{Time} &
\multicolumn{4}{c}{Intensity} &
\multicolumn{1}{c}{Width}  \\ \cline{3-4} \cline{6-8} \cline{10-12} \cline{15-18}
\multicolumn{1}{c}{NOAA} &
\multicolumn{1}{c}{Time [UT]} &
\multicolumn{1}{c}{\#} &
\multicolumn{1}{c}{Code} &
\multicolumn{1}{c}{} &
\multicolumn{1}{c}{PFE} &
\multicolumn{1}{c}{VCA} &
\multicolumn{1}{c}{NLFF} &
\multicolumn{1}{c}{} &
\multicolumn{1}{c}{PFE} &
\multicolumn{1}{c}{VCA} &
\multicolumn{1}{c}{NLFF} &
\multicolumn{1}{c}{} &
\multicolumn{1}{c}{[UT]} &
\multicolumn{2}{c}{[\tablenotemark{a}]} &
\multicolumn{2}{c}{[\tablenotemark{b}]} &
\multicolumn{1}{c}{[pixels]}
}
\decimals
\startdata
11158 &   2011/02/12 15:01:57 &     1 &   AR1-L00 &     &       89 &   \bf{114} &   \underline{138} &     &          74 &     \bf{51} &   \underline{69} &     &   15:26:14 &      60 &         16.0 &   2.66 \\
      &                       &     2 &   AR1-L01 &     &   \underline{195} &      116 &   \bf{286} &     &   \underline{64} &          92 &      \bf{35} &     &   15:32:26 &       9 &         12.3 &   0.93 \\
      &                       &     3 &   AR1-L02 &     &      134 &   \underline{110} &   \bf{111} &     &          26 &   \underline{32} &      \bf{28} &     &   15:19:50 &      25 &         20.6 &   1.42 \\
      &                       &     4 &   AR1-L03 &     &   \bf{78} &       92 &   \underline{86} &     &     \bf{29} &          22 &   \underline{23} &     &   15:32:14 &       8 &         10.6 &   1.36 \\
      &                       &     5 &   AR1-L04 &     &   \bf{101} &   \underline{60} &        78 &     &     \bf{81} &   \underline{199} &          211 &     &   15:31:02 &      26 &          8.7 &   2.44 \\
11190 &   2011/04/15 00:47:05 &     6 &   AR2-L01 &     &      324 &   \underline{\bf{380}} &       860 &     &          66 &   \underline{\bf{55}} &           52 &     &   01:15:02 &     259 &        223.2 &   1.40 \\
      &                       &     7 &   AR2-L02 &     &      205 &      208 &   \underline{\bf{310}} &     &         128 &         115 &   \underline{\bf{121}} &     &   01:17:50 &      12 &         19.0 &   1.02 \\
      &                       &     8 &   AR2-L03 &     &      264 &   \bf{256} &   \underline{244} &     &          48 &     \bf{30} &   \underline{95} &     &   01:11:38 &      16 &         15.0 &   1.34 \\
      &                       &     9 &   AR2-L05 &     &      198 &      143 &   \underline{\bf{173}} &     &         147 &         185 &   \underline{\bf{199}} &     &   01:19:26 &      17 &         18.7 &   1.34 \\
      &                       &    10 &   AR2-L06 &     &   \bf{74} &       66 &   \underline{132} &     &    \bf{144} &         428 &   \underline{164} &     &   01:18:14 &      10 &          5.9 &   1.67 \\
      &                       &    11 &   AR2-L07 &     &   \bf{133} &      147 &   \underline{90} &     &     \bf{83} &          91 &   \underline{325} &     &   01:22:38 &      11 &          8.2 &   1.62 \\
      &                       &    12 &   AR2-L11 &     &      113 &       97 &   \underline{\bf{160}} &     &         130 &         137 &   \underline{\bf{162}} &     &   01:13:38 &      37 &         18.8 &   1.87 \\
      &                       &    13 &   AR2-L13 &     &   \underline{\bf{194}} &      172 &       203 &     &   \underline{\bf{51}} &          54 &           99 &     &   01:17:26 &       6 &         12.8 &   1.12 \\
      &                       &    14 &   AR2-L14 &     &      120 &   \bf{109} &   \underline{137} &     &         116 &    \bf{111} &   \underline{113} &     &   01:18:26 &      14 &         15.2 &   1.23 \\
      &                       &    15 &   AR2-L16 &     &   \bf{345} &      309 &   \underline{581} &     &     \bf{20} &          15 &   \underline{110} &     &   01:18:38 &       7 &         21.8 &   0.79 \\
      &                       &    16 &   AR2-L17 &     &      128 &   \bf{162} &   \underline{195} &     &          61 &     \bf{76} &   \underline{101} &     &   01:17:26 &       6 &          9.3 &   1.07 \\
      &                       &    17 &   AR2-L23 &     &      352 &      347 &   \underline{\bf{476}} &     &          33 &          33 &   \underline{\bf{41}} &     &   01:16:38 &      34 &         31.3 &   1.47 \\
      &                       &    18 &   AR2-L26 &     &      298 &   \underline{\bf{226}} &        59 &     &          38 &   \underline{\bf{70}} &          384 &     &   01:10:14 &     251 &        175.6 &   1.55 \\
      &                       &    19 &   AR2-L28 &     &      259 &      253 &   \underline{\bf{330}} &     &          31 &          25 &   \underline{\bf{53}} &     &   01:14:50 &      59 &         56.0 &   1.32 \\
      &                       &    20 &   AR2-L29 &     &   \bf{329} &      287 &   \underline{458} &     &     \bf{31} &          22 &   \underline{28} &     &   01:17:50 &      19 &         42.5 &   0.87 \\
11339 &   2011/11/08 18:44:43 &    21 &   AR3-L00 &     &      209 &      177 &   \underline{\bf{201}} &     &          61 &          47 &   \underline{\bf{54}} &     &   19:14:50 &      74 &         37.1 &   1.86 \\
      &                       &    22 &   AR3-L01 &     &      203 &      294 &   \underline{\bf{208}} &     &          86 &         109 &   \underline{\bf{37}} &     &   19:16:02 &      85 &         58.0 &   1.57 \\
      &                       &    23 &   AR3-L02 &     &      311 &   \bf{416} &   \underline{523} &     &         113 &     \bf{64} &   \underline{90} &     &   19:14:26 &      20 &         21.0 &   1.30 \\
      &                       &    24 &   AR3-L05 &     &      286 &      198 &   \underline{\bf{235}} &     &         120 &         229 &   \underline{\bf{42}} &     &   19:14:02 &      18 &         15.7 &   1.39 \\
      &                       &    25 &   AR3-L07 &     &   \underline{\bf{154}} &      116 &       143 &     &   \underline{\bf{289}} &         355 &          320 &     &   19:13:02 &      11 &          7.8 &   1.43 \\
      &                       &    26 &   AR3-L08 &     &   \underline{382} &      275 &   \bf{338} &     &   \underline{38} &          45 &      \bf{40} &     &   19:12:02 &      15 &         20.4 &   1.10 \\
      &                       &    27 &   AR3-L11 &     &   \bf{138} &       86 &   \underline{158} &     &    \bf{181} &         462 &   \underline{290} &     &   19:14:14 &       8 &          5.7 &   1.46 \\
      &                       &    28 &   AR3-L15 &     &      423 &   \bf{433} &   \underline{491} &     &          72 &     \bf{48} &   \underline{76} &     &   19:08:50 &      12 &         12.9 &   1.25 \\
      &                       &    29 &   AR3-L16 &     &      173 &      181 &   \underline{\bf{189}} &     &          67 &          86 &   \underline{\bf{184}} &     &   19:14:50 &       5 &          9.9 &   0.92 \\
      &                       &    30 &   AR3-L19 &     &      142 &      107 &   \underline{\bf{152}} &     &         284 &         342 &   \underline{\bf{258}} &     &   19:14:26 &      16 &          4.7 &   2.27 \\
      &                       &    31 &   AR3-L23 &     &   \underline{348} &      409 &   \bf{364} &     &   \underline{36} &          66 &      \bf{35} &     &   19:14:26 &      52 &         19.9 &   2.09 \\
      &                       &    32 &   AR3-L24 &     &   \bf{172} &      127 &   \underline{219} &     &    \bf{122} &         277 &   \underline{118} &     &   19:14:02 &      19 &          6.4 &   2.24 \\
      &                       &    33 &   AR3-L27 &     &      277 &      309 &   \underline{\bf{275}} &     &         139 &          92 &   \underline{\bf{172}} &     &   19:12:26 &       4 &          4.4 &   1.29 \\
      &                       &    34 &   AR3-L32 &     &   \bf{99} &       84 &   \underline{163} &     &    \bf{177} &         515 &   \underline{254} &     &   19:14:50 &      23 &          4.5 &   3.13 \\
\enddata
\tablenotetext{a}{DN $\rm s^{-1}pixel^{-1}$}
\tablenotetext{b}{DN $\rm s^{-1}pixel^{-1}Mm^{-2}$}
\end{deluxetable*}


\begin{figure*}[!htp]
  \includegraphics[bb=15 0 675 113,width=\textwidth]{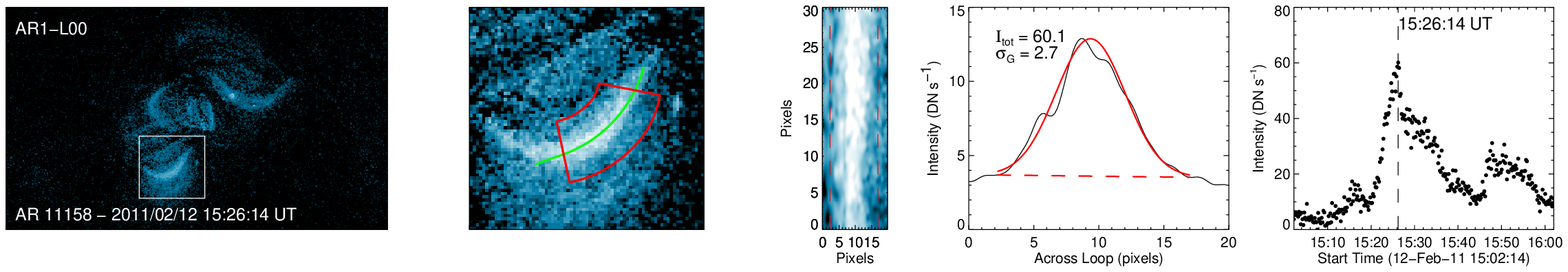}
  \includegraphics[bb=15 0 675 113,width=\textwidth]{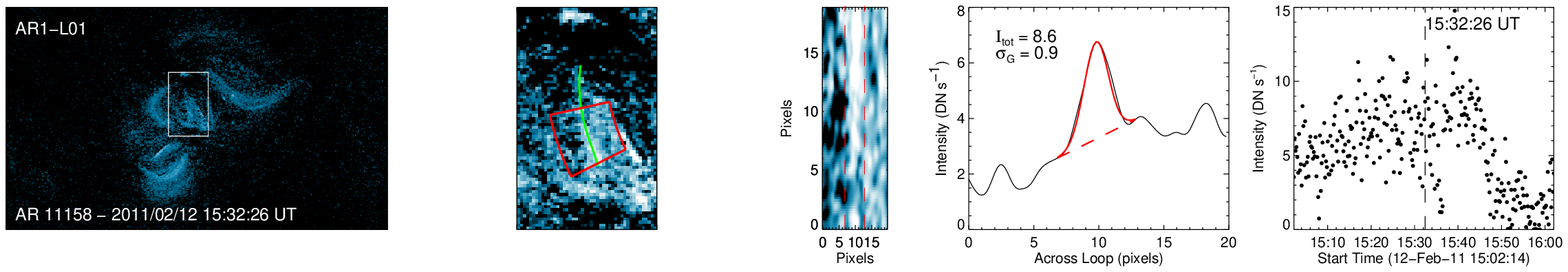}
  \includegraphics[bb=15 0 675 113,width=\textwidth]{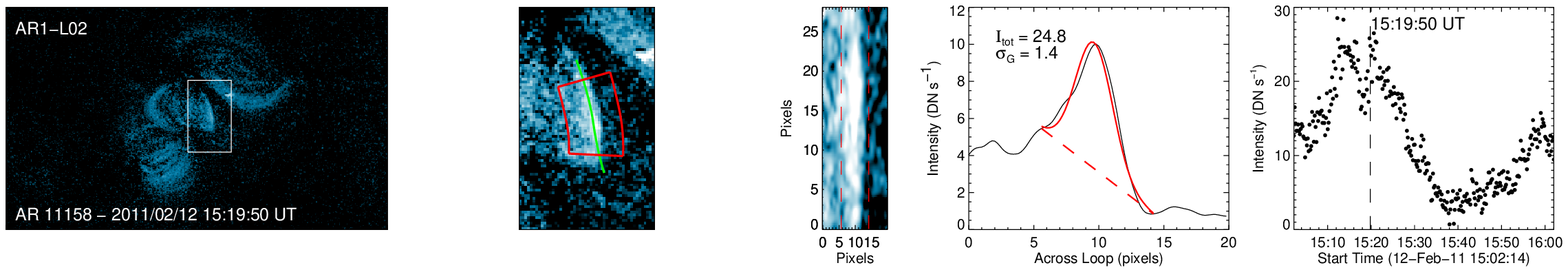}
  \includegraphics[bb=15 0 675 113,width=\textwidth]{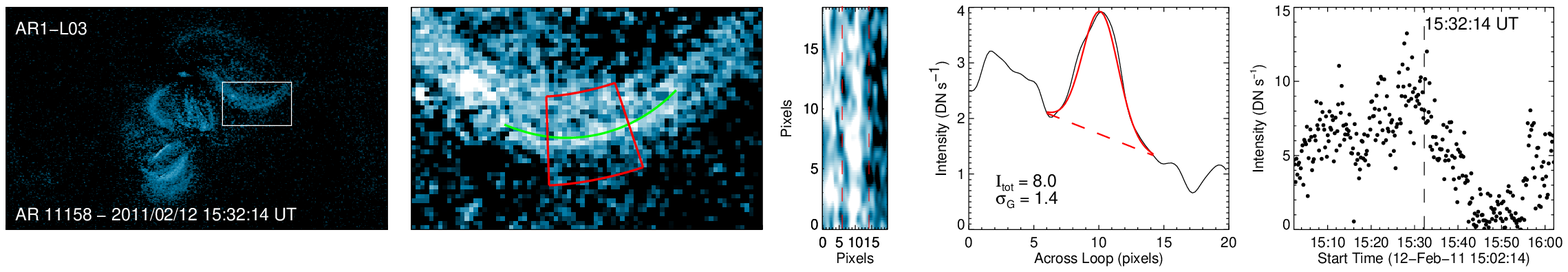}
  \includegraphics[bb=15 0 675 113,width=\textwidth]{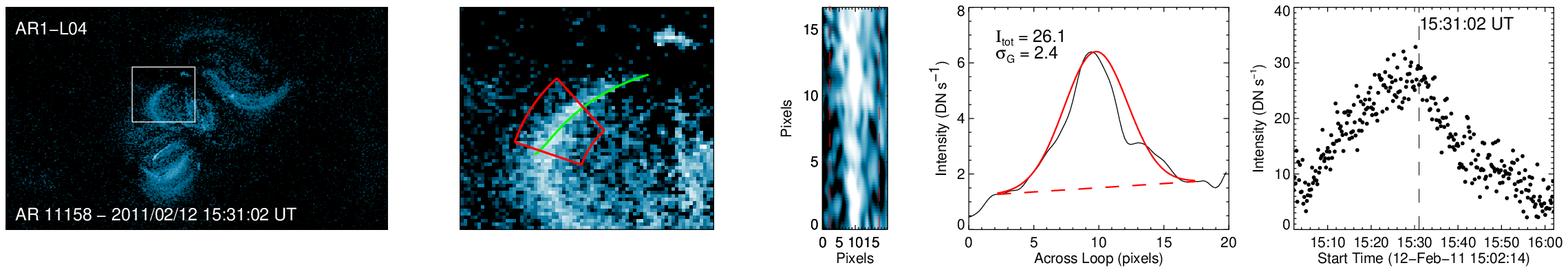}
  \caption{Summary of the analysis performed on the loop segments in active region 11158.
  Labels in the images can be used to link to the loop properties in Table~\ref{tab:loops}. The
  left panels show a wide angle and close-up look at the AIA  \ion{Fe}{18} images of each loop.
  Cross-sectional loop intensities along the loop axis are extracted to form a straightened image in a
  reduced field-of-fiew (red box). The averaged profile is then fitted with a Gaussian (solid red line)
  over a linear background (red dashed line). Finally, the integrated intensity is plotted as a function
  of time. The local intensity peak closest in time to the time of identification (middle of the sequence)
  is saved for comparison to the magnetic properties. The solid green line outlines the location of
  automatic loop segment identification used in the magnetic topology analysis. }
\label{fig:loops_ar1}
\end{figure*}

\begin{figure*}[!htp]
  \includegraphics[bb=15 0 675 113,width=\textwidth]{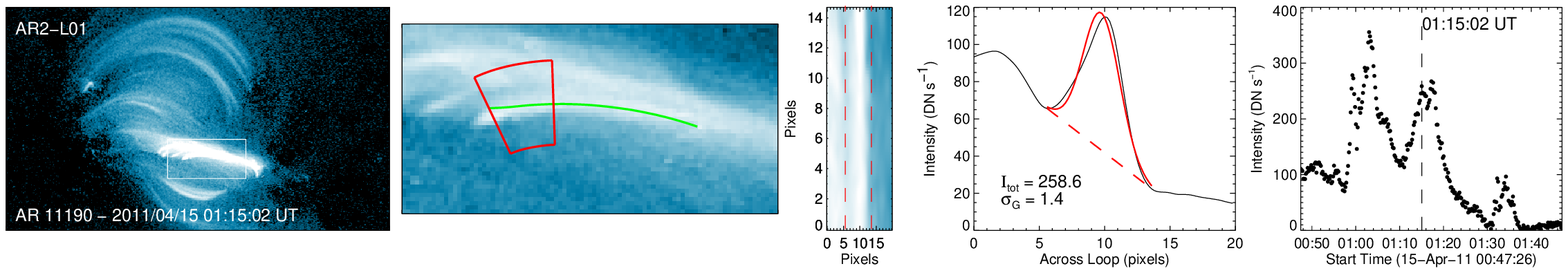}
  \includegraphics[bb=15 0 675 113,width=\textwidth]{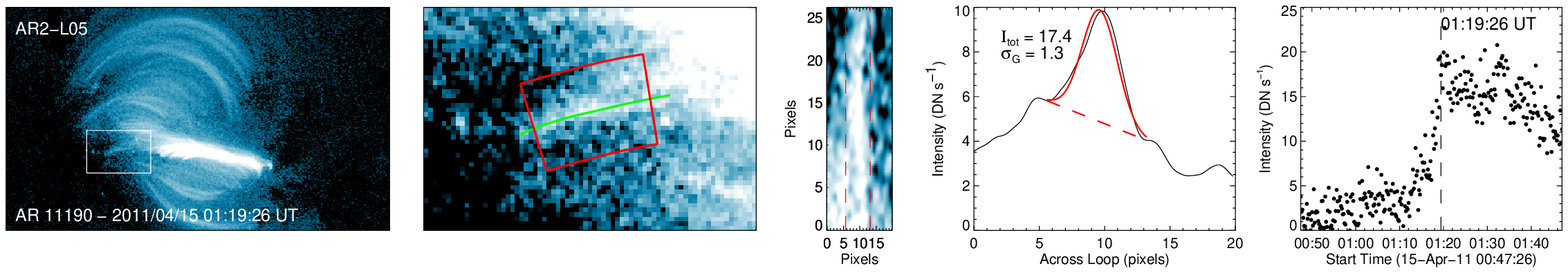}
  \includegraphics[bb=15 0 675 113,width=\textwidth]{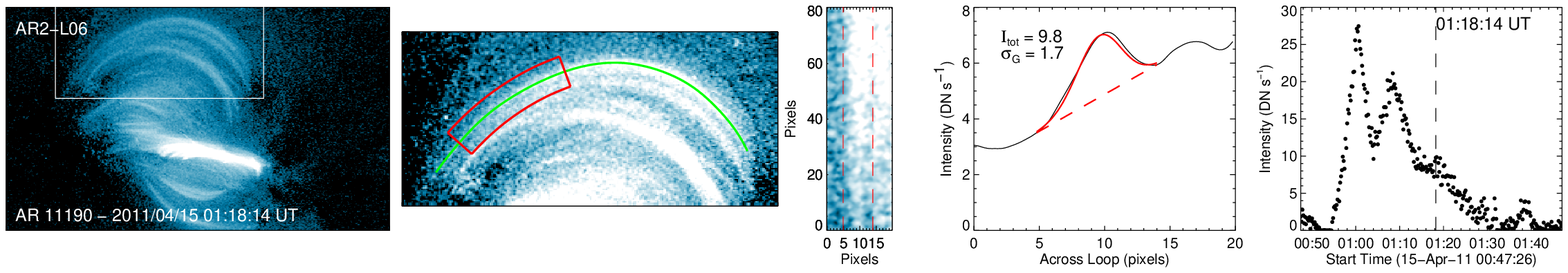}
  \includegraphics[bb=15 0 675 113,width=\textwidth]{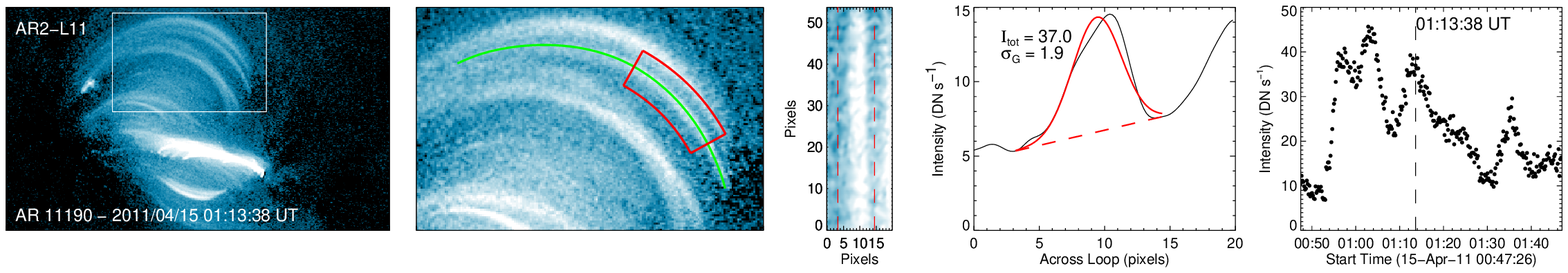}
  \includegraphics[bb=15 0 675 113,width=\textwidth]{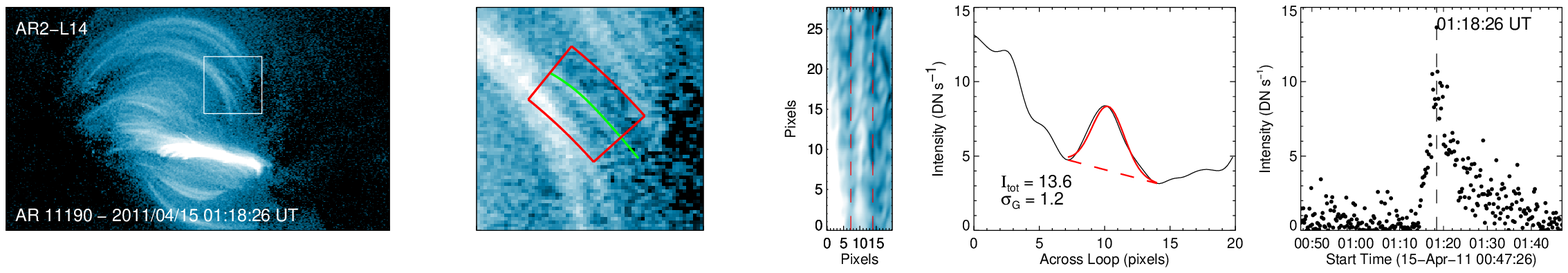}
  \includegraphics[bb=15 0 675 113,width=\textwidth]{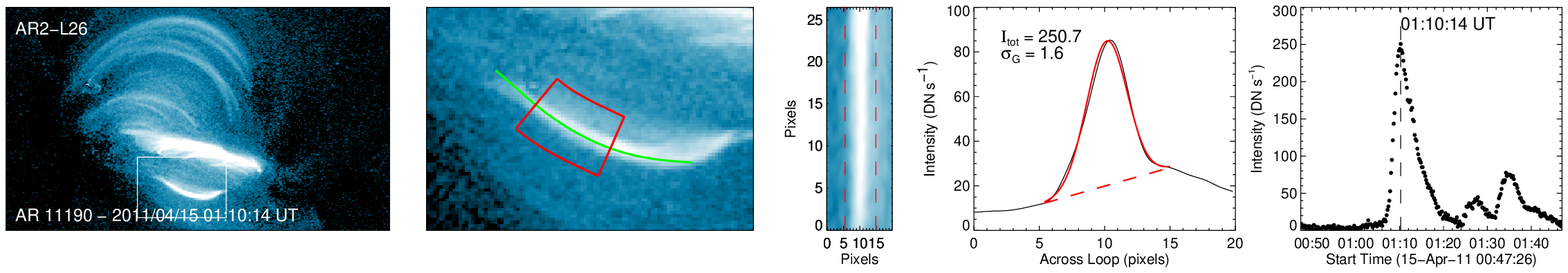}
  \includegraphics[bb=15 0 675 113,width=\textwidth]{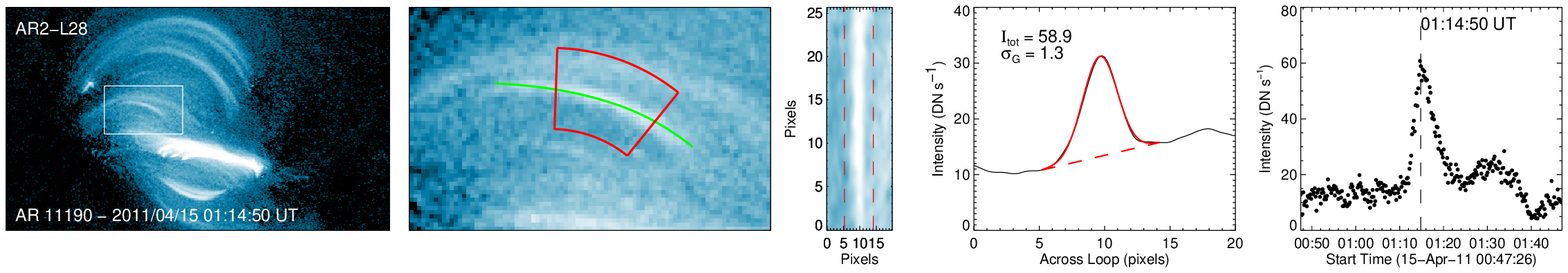}
  \caption{Same as Figure~\ref{fig:loops_ar1} for active region 11190. We show here seven of the fifteen loops studied in this region.}
\label{fig:loops_ar2}
\end{figure*}

\begin{figure*}[!htp]
  \includegraphics[bb=15 0 675 113,width=\textwidth]{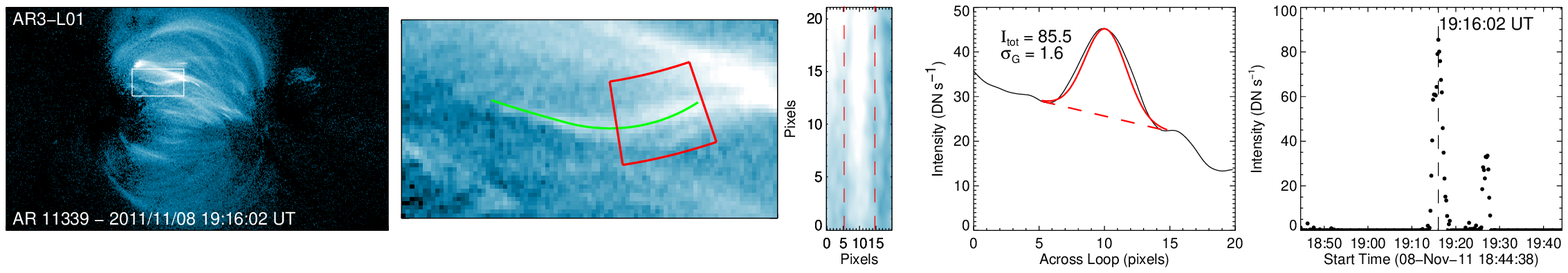}
  \includegraphics[bb=15 0 675 113,width=\textwidth]{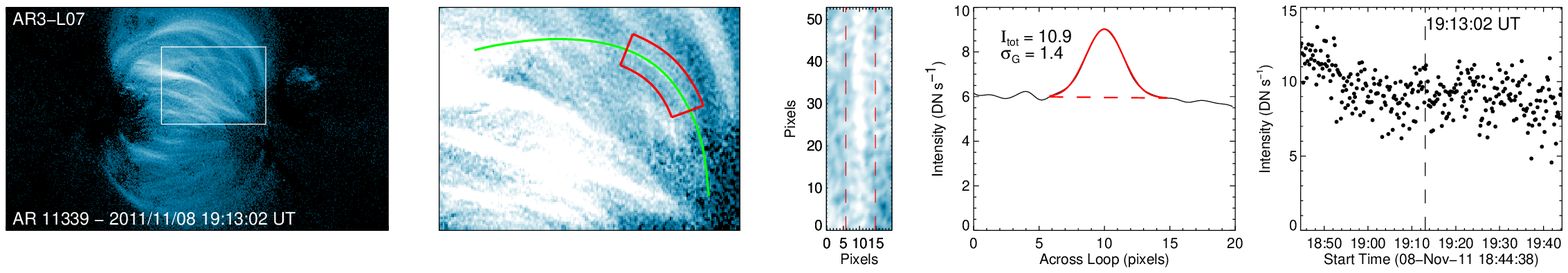}
  \includegraphics[bb=15 0 675 113,width=\textwidth]{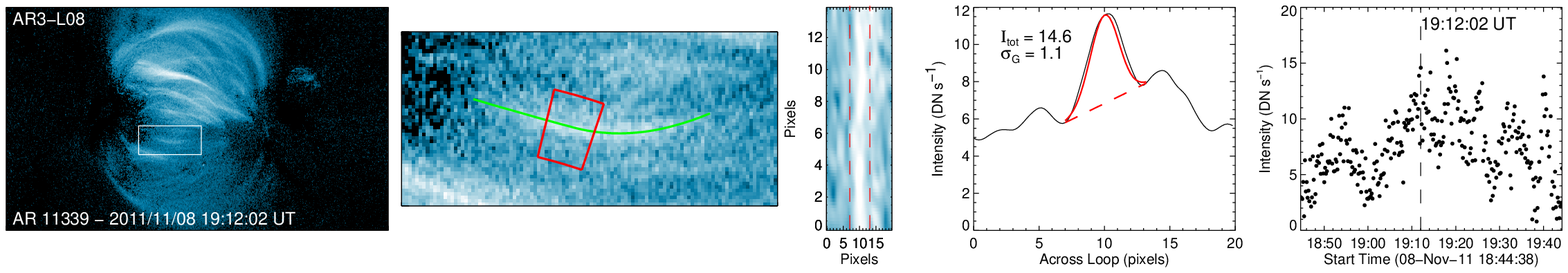}
  \includegraphics[bb=15 0 675 113,width=\textwidth]{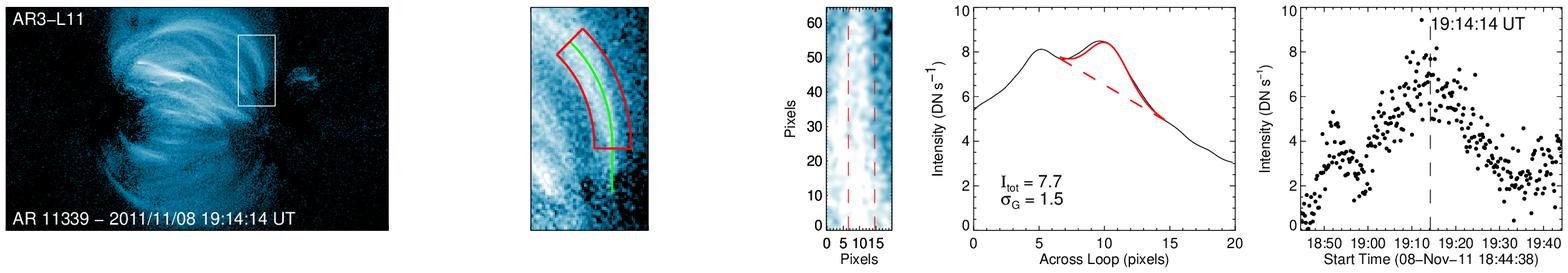}
  \includegraphics[bb=15 0 675 113,width=\textwidth]{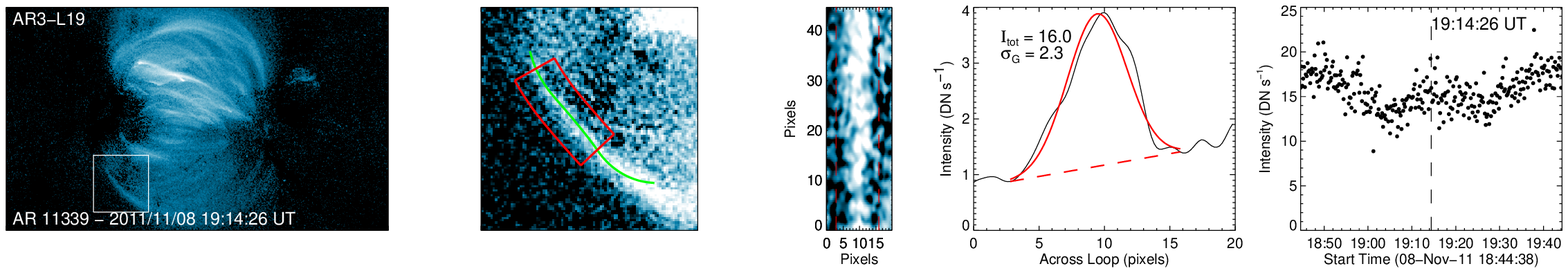}
  \includegraphics[bb=15 0 675 113,width=\textwidth]{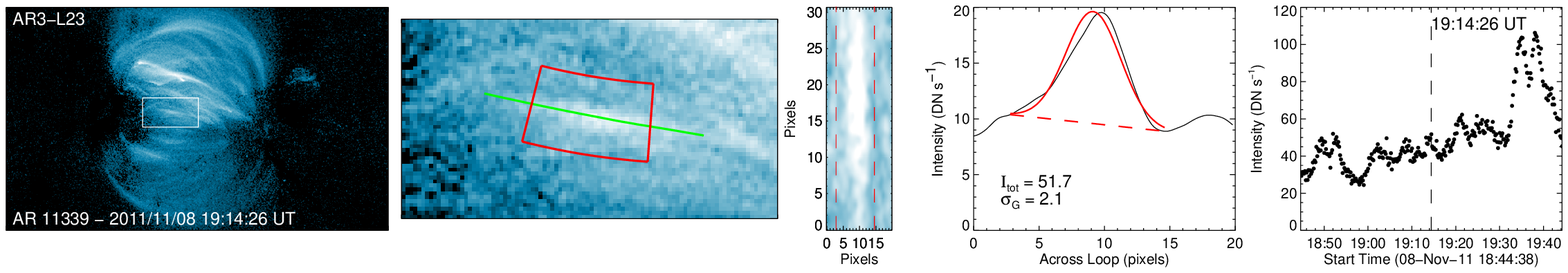}
  \includegraphics[bb=15 0 675 113,width=\textwidth]{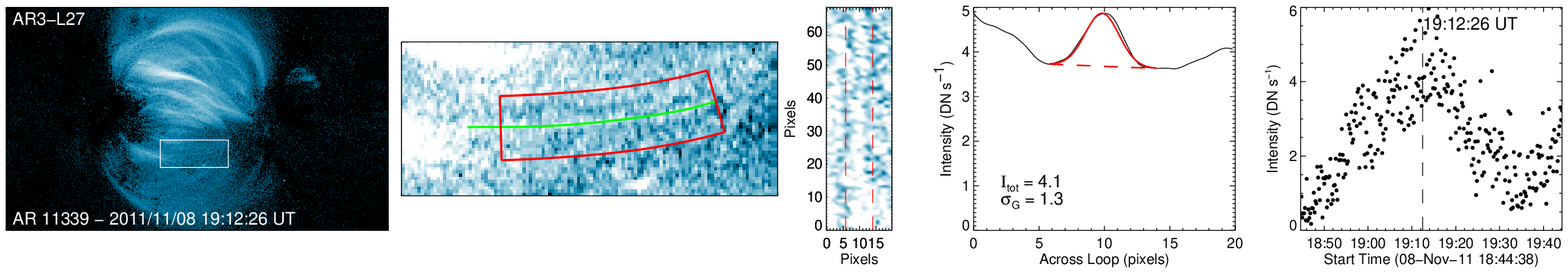}
  \caption{Same as Figure~\ref{fig:loops_ar1} for active region 11339. We show here seven of the fourteen loops studied in this region.}
\label{fig:loops_ar3}
\end{figure*}

\subsection{Mapping field lines to traced loops}

To determine the field line that best matches any given traced loop we used two metrics.
The default metric in our work flow is the ``mean minimum distance'' defined in \citetalias{warren2018}.
This distance is the average minimum Euclidean distance between reference points along the full length
of the traced loop and the field line. This metric works well in most loops, determined by visual inspection.
In 7 of the 34 cases we resorted to a second metric that normalizes the distance along
the traced loop and the field line before evaluating the distance between the points. We developed
this approach for a previous study \citep{ugarte-urra2006} where we found that,  when there are good constraints
on the footpoints, it can help get around misidentifications due to some degeneracy introduced by projecting
3D field lines to an image plane. By comparing proportional distances from the
end points, it helps in discarding longer loops that may project a section of their length
onto the traced loop, resulting in a small Euclidean metric, but an overall incorrect match for
the loop.

We determined the best fits for all identified loops with the two metrics, but only adopted
the ``normalized distance'' metric solution when there were obvious visual cues that this method
provided a better match. The left panel in Figure~\ref{fig:loopmethods}
shows one example for a NLFF extrapolation. In this case the minimum distance method returns a
field line that despite running along most of the traced loop axis, clearly extends
many pixels beyond its observed footpoints. The second metric for this loop returns a smaller
field line that, while it does not provide as good a fit along the axis, we argue that it
provides a better representation of the overall topology (footpoints and projected length)
and therefore a preferred choice to estimate the loop's field strength and length.
Loops that had no satisfactory best-fit field line from either of the metrics were discarded.
This is one of the reasons why the number of loops in the analysis is smaller than the number
of loop segments automatically identified. There were also cases where multiple loop segments
corresponded to the same loop, and we retained the longer ones.

\citetalias{warren2018} provides a systematic comparison between the extrapolation methods,
their ability to reproduce the topology of observed coronal loops, and discusses the best
overall performance. At the individual loop level we find that no single method provides the
best match for all loops. Table~\ref{tab:loops} lists the complete set of loops in this
analysis including the average field strength ($B_{avg}$) and loop length from all three
extrapolation methods. The loops are numbered and have a code identifier that links them to
the active region and the list of loop segments identified for that region. The magnetic properties
for the extrapolation with the smallest minimum distance metric are underlined. The majority
are under the NLFF approach, consistent with \citetalias{warren2018}.
Figure~\ref{fig:loopmethods} shows examples of the different outcomes for the field line
identification from the three methods. We encounter numerous examples where the solutions are
very consistent between the three methods, such as loop \#17 (AR2-L23) in Figure~\ref{fig:loopmethods},
but there are also examples where the discrepancies are significant, e.g. loop \#24 (AR3-L05)
also in the figure. In a few cases, such as loop \#28 (AR3-L15) shown in Figure~\ref{fig:loopmethods},
a visual inspection (space and time) reveals that the loop is best represented by the best-match
field line of an extrapolation (VCA) with a metric value that it is not the smallest (NLFF).
For that reason, we also highlight in Table~\ref{tab:loops} with bold font the field line
selected based on visual inspection of the best-fits for the three methods.
Bold (visual) and underlined (purely metric) selections coincide in fifteen cases and in several others the differences
between the properties are minor. This will become apparent later on when we present the results
as a scatter plot. The inconsistencies ultimately expose the difficulties we faced in disentangling
the projections of a 3D volume of field lines onto the plane of the image. This is particularly
important when loops do not have well identified footpoints and segments cannot impose those
constraints on the field line fitting.

\subsection{EUV Intensities}
\label{sect:intens}

The objective of our study is to investigate the coupling between the loop's magnetic
properties and their radiative output. To characterize the latter, we use the AIA \ion{Fe}{18}
images. The simplest approach would be to record the intensities of the loops at the time
of identification. However, given our hydrodynamic and observational understanding of coronal loop
evolution, this diagnostic seems insufficiently constrained. Loops are often evolving,
changing their brightness, even appearing and disappearing from a particular bandpass
\citep[e.g.][]{winebarger2005,ugarte-urra2009,viall2012}. This behavior is understood
as the signature of changing dynamics in their plasma properties with significant
variations in density and temperature \citep[e.g.][]{reale2014}. A single snapshot is then
likely to portray many loops at different
stages of their evolution. As we intend to compare the intensities for all the loops, we
decided to record the count rates at one particular predetermined instant in their evolution,
the same for all, at the peak in their lightcurves.

The loop intensities are calculated in several steps. We employ an interactive graphical
user interface (GUI) developed by us in IDL. Earlier versions of the tool have been
used extensively to obtain loop widths and intensities in AIA, EIS/{\it Hinode} and
Hi-C data \citep{warren2008,brooks2012,brooks2013}. Figures~\ref{fig:loops_ar1}--\ref{fig:loops_ar3}
show examples of the figures and products generated by the program. The GUI first
imports the location of the automatically detected loop segments to identify the loops
in  the AIA images. This is shown by a green line in the second panel of the figures.
This is only done for the loops that already survived the topological analysis. For every
image in the AIA sequence, the back-end of the program extracts the intensity profiles
across the loop axis, defined here by the imported locations, and along the full length
of the segment, creating an interpolated straightened image of the loop as shown in the
third panel of the figures.
The length of this new loop segment is sometimes smaller than the original outline. This
resizing of the areas of interest (red box in the figures) was necessary to avoid as much
as possible contamination from neighboring loops along the full temporal sequence. The
GUI shows the evolution in time of the AIA images with respect to those choices and allows
for interactive changes.
Intensity profiles are averaged along the loop axis and two background points are
manually selected as part of a Gaussian fit to the loop cross-section. Finally,
the integrated Gaussian intensity for every image is computed and
plotted as a function of time (right-most panel in the figures).

\begin{figure*}[t!]
  \includegraphics[bb=10 40 315 280, clip=true, width=0.33\textwidth]{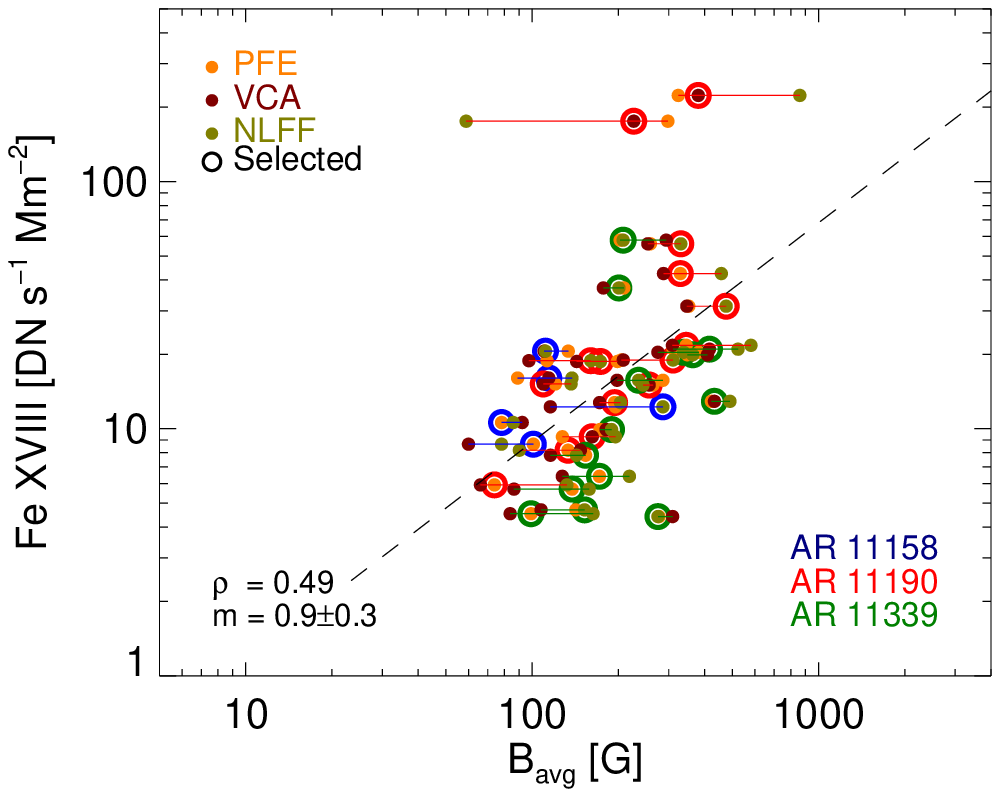}
  \includegraphics[bb=10 40 315 280, clip=true, width=0.33\textwidth]{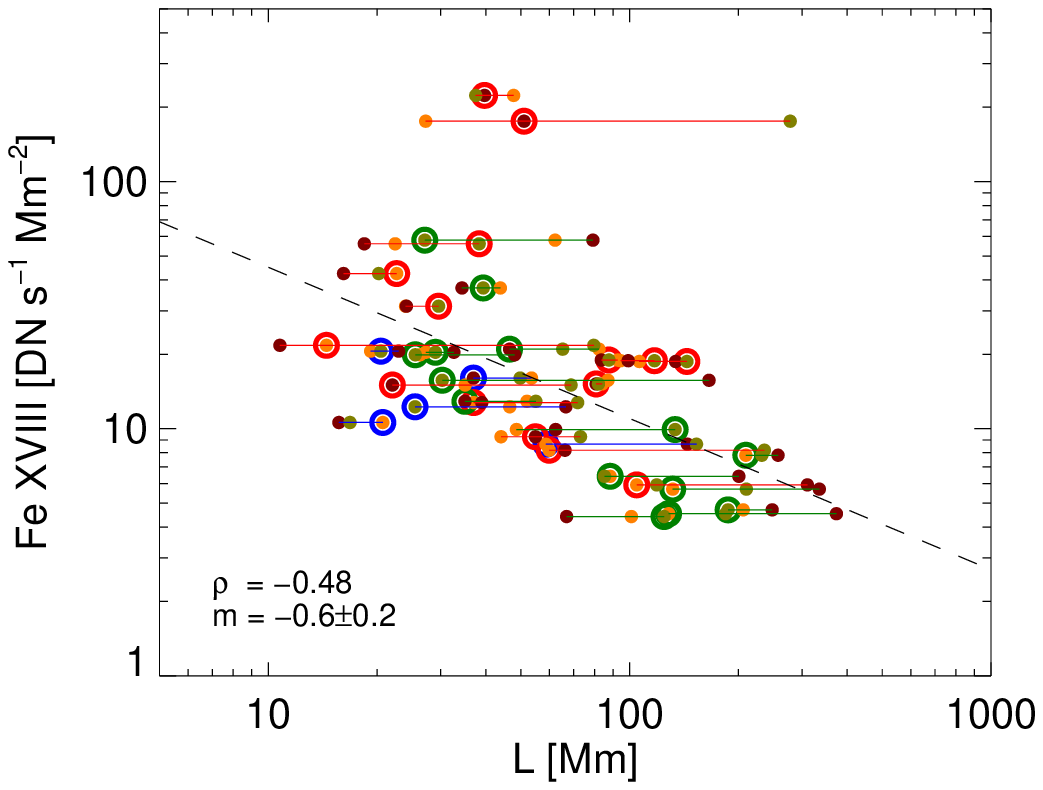}
  \includegraphics[bb=10 40 315 280, clip=true, width=0.33\textwidth]{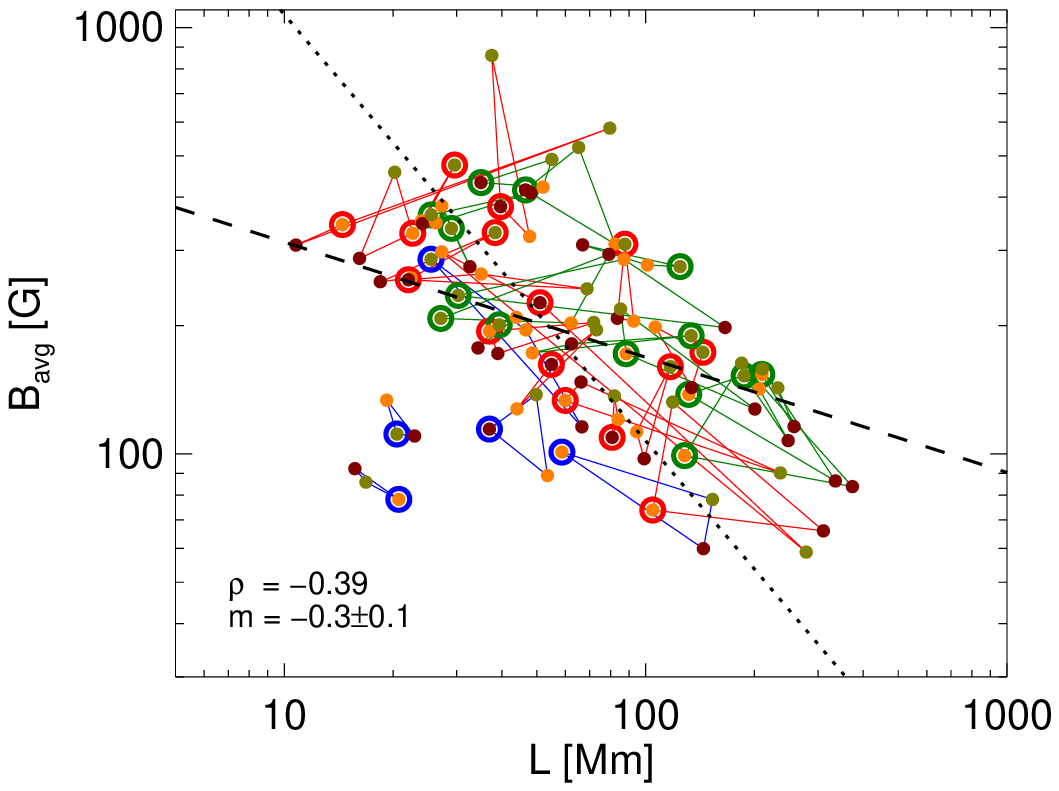}
  \caption{Relationship between \ion{Fe}{18} intensity, average field strength ($B_{avg}$) and loop length ($L$) for
  all 34 loops in the sample presented in Table~\ref{tab:loops}. Filled circles in different colors indicate the
  extrapolation method used in computing the values. Solutions for a single loop are joined by a solid line.
  Open circles highlight the solution, out of the three, selected from visual inspection of the best fitted field lines.
  Colors of open circles and solid lines indicate the active region to which the loop belongs to. Correlation coefficient ($\rho$) and slope ($m$) of
  the power-law fit to the open circles are provided on the legend. The dotted line on the right panel corresponds to
  a slope of $m=-1$ and it is provided for context.}
\label{fig:intvsBavg}
\end{figure*}

We performed this analysis for all loop segments. Several loops had to be discarded
because they were too weak to provide an averaged loop cross-section that could be
fitted. Figures~\ref{fig:loops_ar1}--\ref{fig:loops_ar3} show representative cases
for all three regions. Table~\ref{tab:loops} shows the resulting intensities and widths for
all the loops. The intensities correspond to the local peak closest in time to the
middle of the sequence, the time where the automatic loop identification algorithm
was applied. As the emission in coronal plasmas depends on the volume of the emitting
structure, to compare the intrinsic emission from loops with different integration
paths we divide the count rates by the cross-sectional area. We assume circular
cross-sections with a diameter that is twice the Gaussian standard deviation. This is presented
in Table~\ref{tab:loops} in units of DN s$^{-1}$ Mm$^{-2}$.

Most loops exhibit lightcurves characteristic of transient events: a rising phase,
followed by a peak and a decaying phase within the short observing window. There are,
however, a few cases such as AR3-L07 (Figure~\ref{fig:loops_ar3}), where the lightcurve
remains mostly constant. Long-lived loops have been observed before
\citep[e.g][]{lopezfuentes2007,scott2012}, particularly at high temperatures.
In our dataset, transients are more common.

Our ability to subtract the background intensity is one of the main sources of
uncertainty in measuring the intensities of the loops. The spread of the data points
in time in the lightcurves gives us an estimate of the uncertainties in the measurement,
which goes from $50\%$ in a weak loop (AR3-L27) to $2\%$ in a bright one (AR2-L26).

\section{Results}
\label{sect:results}
In this section we describe how the various properties of the loops
(radiative, magnetic, geometrical), obtained through the methods presented earlier,
scale with each other.

The left panel in Figure~\ref{fig:intvsBavg} shows how the peak in \ion{Fe}{18} intensity
scales with the average magnetic field strength ($B_{avg}$) for all loops in the three active
regions. In the figure, each loop has three $B_{avg}$ values, joined by
a solid line, corresponding to three extrapolation methods presented in Table~\ref{tab:loops}.
This provides an impression on the level of precision we achieve in determining this quantity
using three standard models to describe the magnetic properties of loops in the corona.
There are no systematic differences between the three,
which suggests that the results of this study are largely independent of the magnetic model.
Note, however, that we have not tested the accuracy of the models.
The large open circles highlight
our preferred solution from visual inspection of the three best-matched field lines, i.e.
the values in bold font in Table~\ref{tab:loops}. We use only these points in any subsequent
calculations. We find that there is a correlation of $\rho=0.49$ between the \ion{Fe}{18}
intensity and $B_{avg}$. A linear fit to the selected points returns a power-law relationship with
a slope of $m=0.9\pm0.3$. As we average the counts from many pixels along and across every
loop, systematic error dominate over the statistical uncertainties in the intensities. The
fit parameters are reproducible when we fit a subset of the data points with a Monte Carlo approach.
We obtained the same slope and standard deviation from the histogram of solutions to fits of 100
random combinations of 25 data points.

The middle panel shows how the \ion{Fe}{18}  intensity scales with loop length. They are
anticorrelated ($\rho=-0.48$) which confirms the general visual impression from coronal images
that the longer the loops are, the fainter they emit at these wavelengths.

Two data points at large intensities (AR2-L01, AR2-L26) appear outliers to the general
trend in both panels. We can not explain their excess of counts (factor of 4) by uncertainties
in the fits to the loop intensities. The signal is high and the profiles are clean. One
possibility is that we have underestimated the volume of emission by assuming a circular
cross-section. Multiple loops brightened up at the same time nearby, and while we removed the
background, an alignment along the line-of-sight can not be ruled out. The other possibility is
that the change in behavior is real and needs a physical explanation. As we do not have a reason
to discard them, we keep the two points as part of the analysis.

The right panel in Figure~\ref{fig:intvsBavg} shows the relationship between field strength and loop length. This relatively
complex figure shows the same information as previous panels, but now the three topological
solutions for a particular loop constitute the three vertices of a triangle joined by a solid
line. The closer the vertices, the better agreement between the methods. The spread of the
solutions shows that, despite the general agreement in the dependencies of the intensity with
$B_{avg}$ and $L$, there can be significant differences between the [$B_{avg}$, $L$] pairs
for each method.
The correlation coefficient is $\rho=-0.39$ for the overall set of points with a slope
of $m=0.3\pm0.1$. For reference, \citet{mandrini2000} found
slopes of $-0.97\pm0.25$ using all field lines in their extrapolations within predetermined
thresholds, but not isolating the ones linked to emitting loops. The dotted line shows a
slope of -1 for context. The loops in our sample are not a distinct population within the
general distribution of $B_{avg}$ and $L$, except for having average field strengths that are
consistently larger than $\sim$100 G.

\begin{figure}[t!]
  \includegraphics[bb=10 75 330 280,clip=true, width=0.48\textwidth]{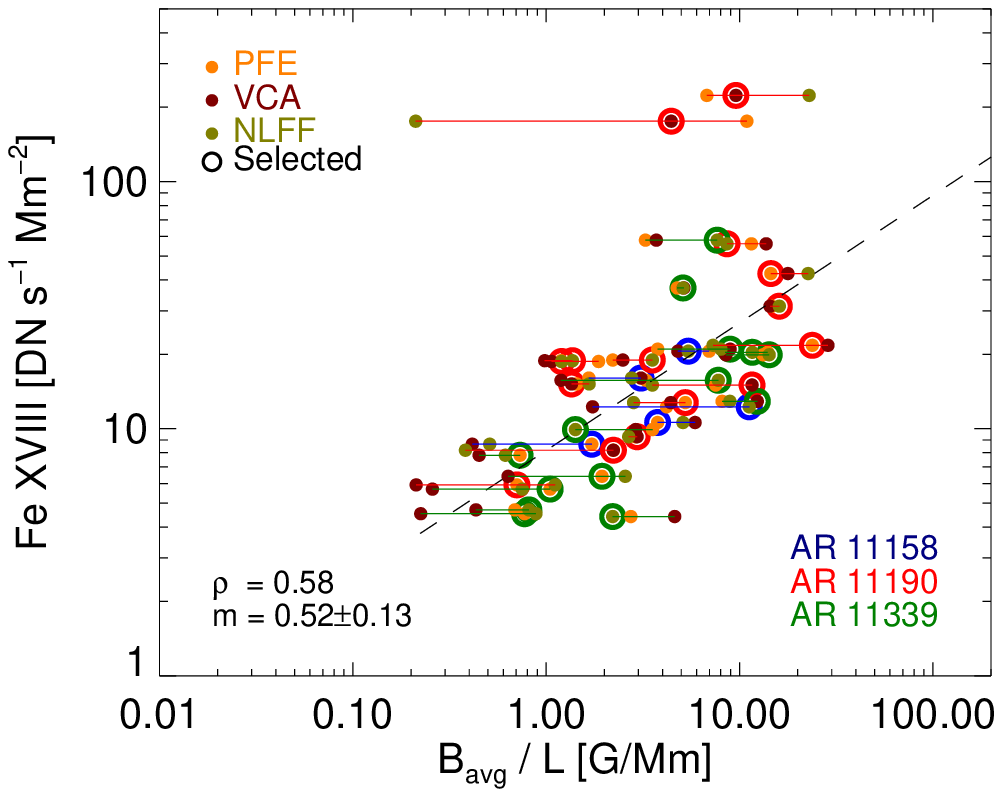}
  \includegraphics[bb=10 40 330 280,clip=true, width=0.48\textwidth]{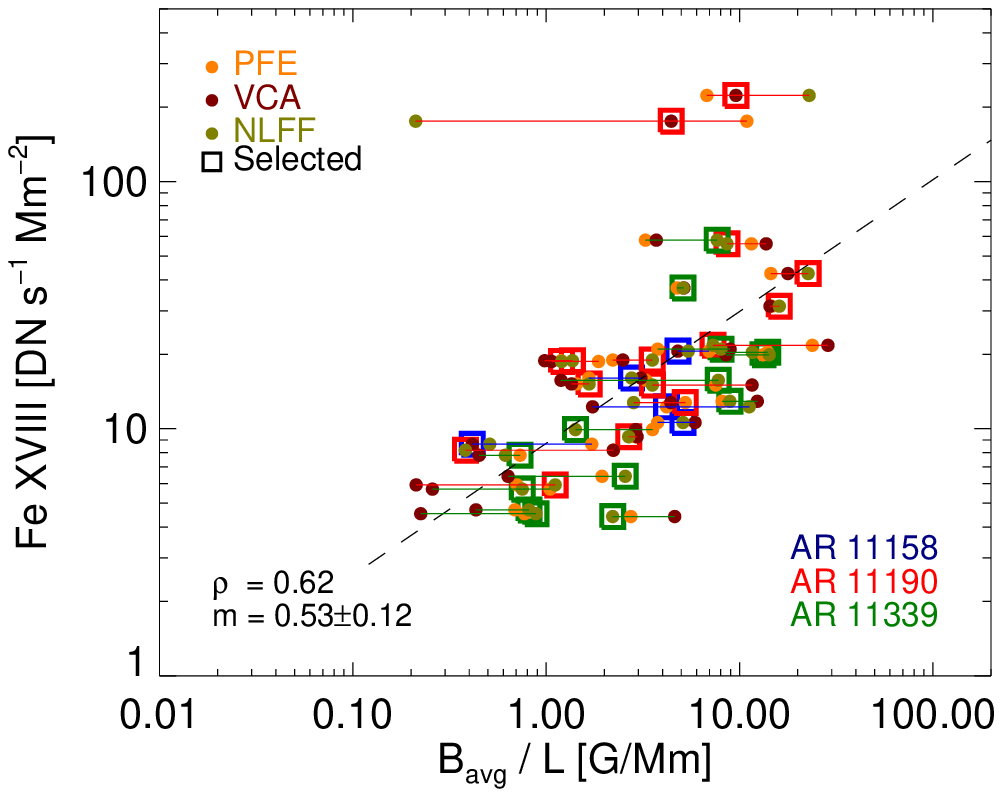}
  \caption{Relationship between \ion{Fe}{18} intensity and the $B_{avg}/L$ ratio  for
  all 34 loops in the sample. Symbols and colors are the same as in Figure~\ref{fig:intvsBavg}.
  The top panel highlights with open circles the solution selected, out of the three
  extrapolation methods, from visual inspection of the best fitted field lines  (bold
  font values in Table~\ref{tab:loops}). The bottom panel highlights with open squares
  the solution with the smallest metric (underlined values in Table~\ref{tab:loops}).}
\label{fig:intvsBavgoverL}
\end{figure}

\begin{deluxetable}{lDD}
\tabletypesize{\small}
\tablecaption{Correlation coefficient and slope of the power-law fit
to the \ion{Fe}{18} intensity dependence on the following quantities \label{tab:correlations}}
\tablewidth{0pt}
\tablehead{
\multicolumn{1}{c}{} &
\multicolumn{2}{c}{$\rho$} &
\multicolumn{2}{c}{$m$}
}
\decimals
\startdata
$L$                & $-0.48$ & $-0.6\pm0.2$  \\
$B_{avg}$          & $0.49$  & $0.9\pm0.3$   \\
$B_{apex}$         & $0.55$  & $0.7\pm0.2$ \\
$B_{f}$    & $0.02$  & $0.0\pm0.3$   \\
$B_{fmax}$    & $0.02$  & $0.0\pm0.3$   \\
$B_{fmin}$    & $0.18$  & $0.3\pm0.3$   \\
${B_{avg}/L}$      & $0.58$  & $0.5\pm0.1$ \\
$B_{apex}/L$       & $0.57$  & $0.4\pm0.1$ \\
$B_{f}/L$  & $0.50$  & $0.6\pm0.2$   \\
$B_{fmax}/L$  & $0.47$  & $0.6\pm0.2$   \\
$B_{fmin}/L$  & $0.52$  & $0.5\pm0.2$ \\
\enddata
\end{deluxetable}


In Figure~\ref{fig:intvsBavgoverL} we show the dependence of \ion{Fe}{18} intensity
as a function of the $B_{avg}/L$ ratio. The top panel shows the fits for the selection
of best visual solutions as in Figure~\ref{fig:intvsBavg}. The correlation clearly
increases with the combined dependence ($\rho=0.58$). As in the previous panels the
linear fit ($m=0.52\pm0.13$) is not satisfactory for all data points, but we prefer
to leave any non-linear description to future samples with more points on the the high
intensity end. In the bottom panel, we show the same exact figure but highlighting with
open squares the solutions with the absolute smallest distance metric. The results overlap and
demonstrate that the trends are robust and not subject to any bias introduced in the
visual selection.

In Table~\ref{tab:correlations} we show the correlation coefficients and slopes
for these quantities and others that we also considered. We include the magnetic
field strength at the apex of the loop ($B_{apex}$) and at the footpoints: as an
average ($B_f$), the maximum ($B_{fmax}$), and the minimum
($B_{fmin}$). We find the field strength at the footpoints to be uncorrelated
to the \ion{Fe}{18} intensity of the loops. The results for the apex are interesting
because they exhibit the highest correlation for any field strength measurement.
This may, however, be driven by the anticorrelation between $B_{apex}$ and loop length
($\rho=-0.63$). In fact, $B_{apex}/L$ is not a better proxy for \ion{Fe}{18}
intensity than $B_{avg}/L$.

\section{Discussion}
\label{sect:discussion}
There have been many studies
that have shown that the luminosity or the total radiance of active regions and other features
on the Sun scale up with the total unsigned magnetic flux as a power-law relationship
\citep[e.g.][]{schrijver1987,fisher1998,benevolenskaya2002,fludra2002,pevtsov2003,fludra2008}.
Models of these regions as the sum of individual loops with a prescribed volumetric
heating rate of the form $\epsilon_H\propto B^{\alpha} / L^{\beta}$, where $\alpha$ and $\beta$
are positive, have been successful at describing global intensities. The actual
values of those exponents have been a source of debate in the literature:
\citet{schrijver2004,warren2006,warren2007,lundquist2008b,winebarger2008,dudik2011,ugarte-urra2017}.
Recent 3D MHD resistive simulations of magnetic flux tubes shuffled at their footpoints
also predict that loop properties such as the temperature should scale proportionally
with magnetic field strength and to a lesser degree with the inverse of the loop length
\citep{dahlburg2018}.

In that context, the main result of our study, the scaling of the loop's radiance
with magnetic field strength and the inverse of the loop's length is not unexpected.
It is, nonetheless, an independent estimate that had not been established yet. The novelty
stands in that we use measurements of individual loops (intensities, cross-sections),
that are then combined with the still unavoidable magnetic models that previous studies
already used. We link for the first time at these wavelengths the peak emission of what
we interpret as a heating event in an evolving loop structure to its magnetic properties.
A similar loop survey was recently carried out by \citet{xie2017}, who looked at the spectroscopic
and magnetic properties of 50 active region loops with temperatures $\leq2$ MK. While the spectral
analysis allowed them to obtain valuable plasma properties such as velocity and density, that
study lacks the temporal resolution to investigate the evolution in time that we argue here is
critical for our purposes of characterizing the same evolutionary stage for all loops.

These results provide new constraints for future modeling efforts of the corona.
Progress in the field over the past 30 years has meant that simply showing that our favorite
heating mechanism or numerical experiment produces million degree temperatures is no longer
sufficient. A viable model must reproduce the observed relationships between the magnetic
properties of loops and the radiative signatures.
As our results illustrate, the constraints from observations
are significantly more precise than just a target temperature. Recent efforts looking at how
models compare to constraints from emission measure distributions \citep{cargill2014,dahlburg2016,barnes2016},
time-lags \citep{bradshaw2016}, Doppler shifts \citep{hansteen2010,zacharias2011,bourdin2013}
or non-thermal velocities \citep{olluri2015}, are examples of the type of tests that numerical
experiments need to pass before declaring success for any given theory. At this point, no
experiment has been able to model active region heating, reproducing simultaneously a multiple
set of constraints, such as emission measure distributions (i.e. emissivity a various temperatures),
scaling with field strength, short-term variability, long-term decay, detailed morphology
(i.e. topology), loop widths, etc. Significant progress has been made, however, in many of
those areas independently.

Before we delve into the implications of our results for the heating rate, it is worth
stopping to discuss other (non) dependencies. While it was not unexpected to find
the scaling of intensity with $B_{avg}$, the fact that the \ion{Fe}{18} intensity
is uncorrelated with field strength at the footpoints ($B_{f}$) does seem surprising. After
all, the source of the energy in the system is expected to come from the convective motions
at the surface where the feet of the loops are rooted, and there have been several studies
arguing that the energy is preferentially deposited at the lower layers of the atmosphere:
spicules contribution \citep{depontieu2009,depontieu2011}, thermal non-equilibrium
as a result of preferential footpoint heating \citep[e.g.][]{mikic2013,froment2017},
predictions from 3D MHD resistive models \citep[e.g.][]{gudiksen2005,hansteen2010,reale2016}.
See also the broader remarks from \citet{aschwanden2007} and the counter arguments
from \citet{klimchuk2015} and references therein. Our results, where the intensity
correlates with a quantity that depends on the field all along the loop, seem to be
consistent with the coronal heating argument. We cannot, however, underestimate the
limitations of current magnetic models. In fact, the spread of the solutions for each
loop on the right panel of Figure~\ref{fig:intvsBavg} indicates that we still need to
improve our understanding on how accurate the different magnetic models are in
reproducing coronal conditions.

This study would not be complete if we did not discuss to some level
the coronal heating implications of these loop measurements. As emission is only
one by-product of heating in the atmosphere, to infer any of its underlying
properties it is necessary to employ a model that describes how heating turns into
emission at these wavelengths. Assuming that the evolving loops that we are
observing can be modeled as independent single hydrodynamic structures, we
use the "Enthalpy-based Thermal Evolution Loops" (EBTEL) model \citep{klimchuk2008,cargill2012}
to study the response of the plasma to heating. In particular, we are interested in
how the \ion{Fe}{18} intensity scales with a heating that depends on $B_{avg}$ and
$L$. We have looked at a parameterization of the volumetric heating rate that scales as
\begin{equation}
\epsilon_{H} = \epsilon_{0} \left(\frac {B_{avg}} {B_{0} } \right)^{\alpha} \left(\frac{L_{0}}{L} \right)^{\beta}
\end{equation}
similar to our previous studies \citep{warren2006,warren2007,ugarte-urra2017}.
Following \citet{warren2006} we start by assuming $\epsilon_{0}=0.0492\, \rm \,ergs \,cm^{-3} \,s^{-1}$
and $B_0=76 \rm \,G$ and $L_0=29 \rm \,Mm$ for an apex temperature close to 4 MK.
For this work we find that $\epsilon_{0}=0.0738$ is a better choice to match the \ion{Fe}{18}
intensities. That corresponds to a static equilibrium temperature of $\sim$5 MK.

\begin{figure*}[!htp]
  \includegraphics[clip=true, width=\textwidth]{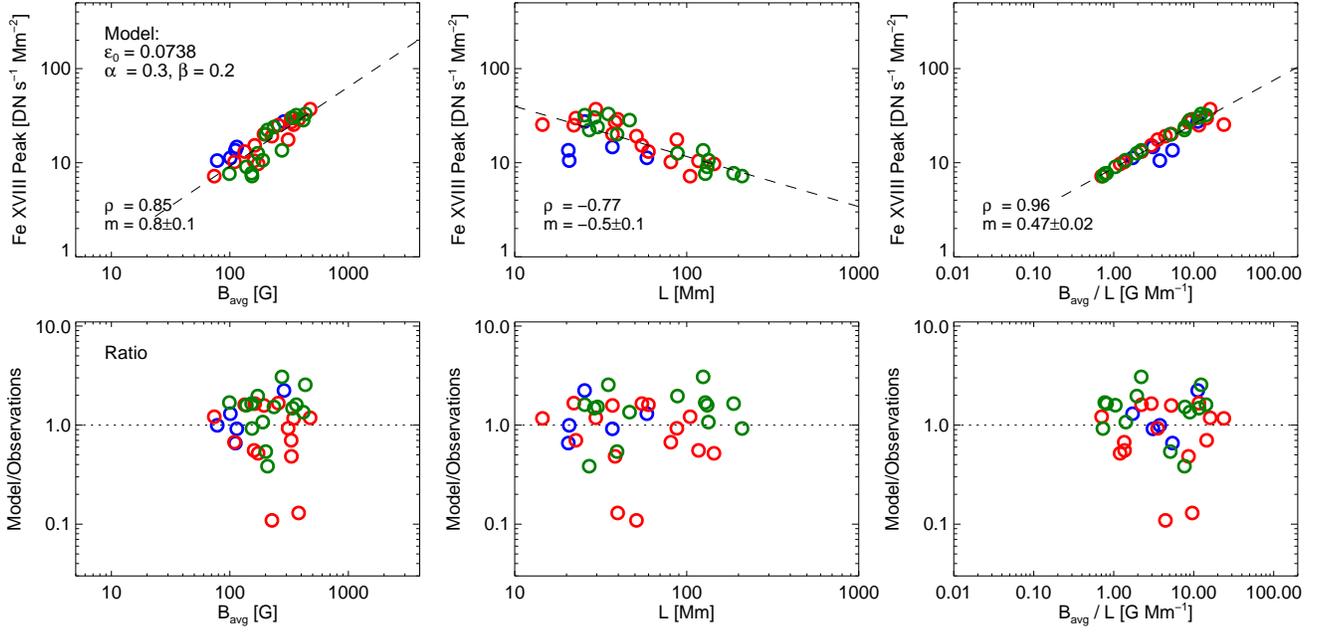}
  \caption{Dependence of the simulated \ion{Fe}{18} peak intensity as a function of $B_{avg}$
  and $L$. The top row shows the results of the model that best matches the observed slopes
  presented in Figures~\ref{fig:intvsBavg} and ~\ref{fig:intvsBavgoverL}. The bottom row shows
  the ratio of model to observed intensities. The colors of the symbols have the same coding
  as previous figures.}
\label{fig:model_best}
\end{figure*}
\begin{figure*}[!hp]
  \includegraphics[clip=true, width=0.96\textwidth]{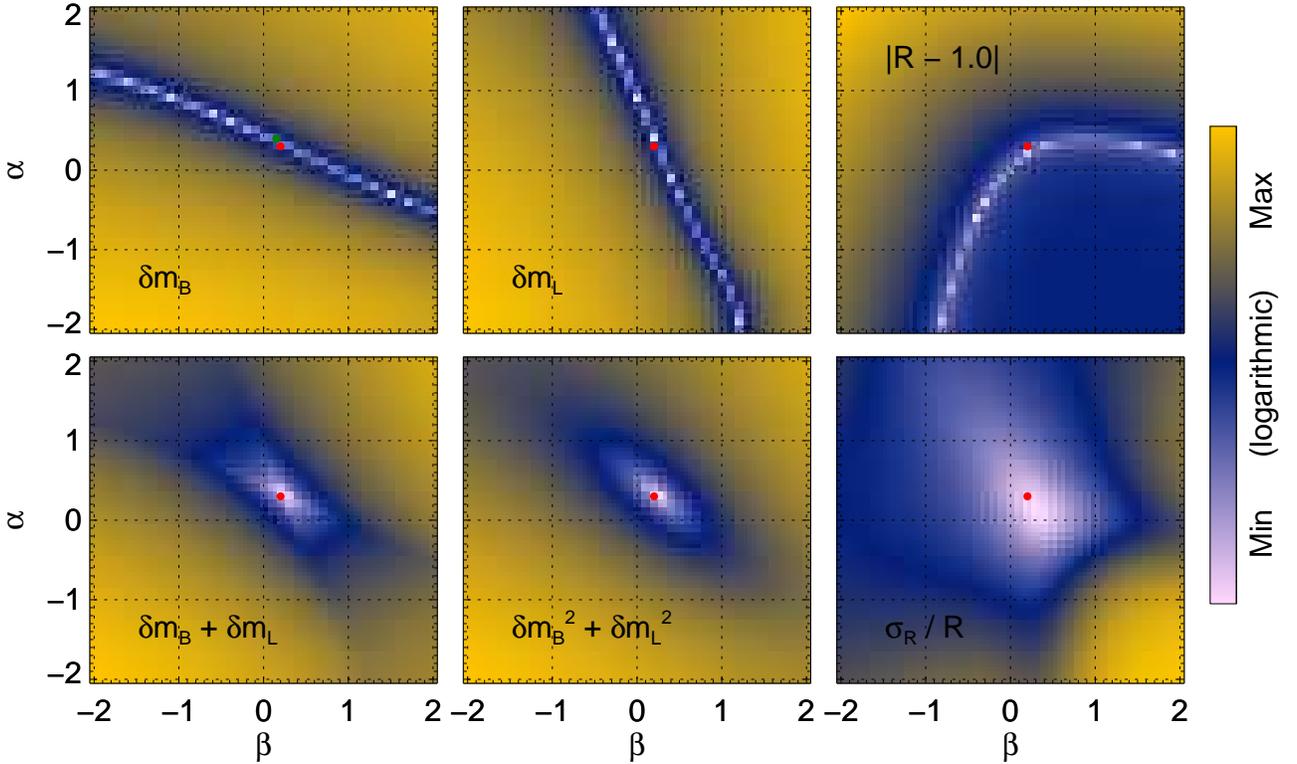}
  \caption{Parameter space explored when modeling impulsively heated loops with a
  volumetric heating rate parameterized as $B_{avg}^\alpha/L^\beta$. The red symbol shows
  the best fit solution ($\alpha=0.3$, $\beta=0.3$) for the combined slopes ($\delta m_B + \delta m_L$
  and $\delta m=|m_{model} - m_{obs}|$)
  of the \ion{Fe}{18} intensity dependence on $B_{avg}$ and $L$. Also shown,
  the average ratio of intensities ($R = I_{model}/I_{obs}$) and standard deviation ($\sigma_{R}$).}
\label{fig:model_paramspace}
\end{figure*}

We computed the hydrodynamic response of all loops in our sample as determined
by their $B_{avg}$ and $L$ as shown in bold font in Table~\ref{tab:loops}.
The heating time profile was assumed as a 200s triangular pulse with a $\epsilon_H$
peak, over a steady background (static equilibrium temperature of 0.5 MK).
From the temporal evolution of density and temperature, assuming 350 km radius loops
\citep{ugarte-urra2017}, and the help of the CHIANTI atomic database \citep{dere1997,delzanna2015},
we computed the AIA \ion{Fe}{18} lightcurves and recorded the peak time and intensity.
For each loop we explored exponents $\alpha$ and $\beta$ ranging [-2,2] in increments
of 0.1. For each pair of exponents we calculated the dependence of the AIA \ion{Fe}{18}
intensity with $B_{avg}$, $L$ and the $B_{avg}/L$ ratio and performed a linear fit to the
power-law.
The heating parameterization that best reproduces the observed slopes of the $B_{avg}$
and $L$ dependence is $\alpha=0.3$ and $\beta=0.2$. Within the uncertainties of the
observed slopes, other pairs of solutions are still compatible, and lie in a diagonal
where increasing $\alpha$ decreases $\beta$ and viceversa. The approximate range is
$\alpha=0.3\pm0.2$ and $\beta=0.2\pm^{0.2}_{0.1}$, with [$\alpha$,$\beta$]=[0.5,0.1] and
[0.1,0.4] at the extremes of the diagonal.
Figure~\ref{fig:model_best}
shows the dependencies and the ratio of modeled to observed intensities for that case.
Figure~\ref{fig:model_paramspace} shows the full
parameter space of residuals resulting from comparing the slopes of the EBTEL modeling
and AIA observations ($\delta m=|m_{model} - m_{obs}|$), with the minimum at $\alpha=0.3$
and $\beta=0.2$ for $\delta m_B + \delta m_L$ and $\delta m_B^2 + \delta m_L^2$. In the
figure we also show the average ratio of intensities ($R = I_{model}/I_{obs}$) and the
standard deviation ($\sigma_{R}$). Both also reach a valley around that solution.

This result compares well with several previous studies. \citet{warren2006} and \citet{lundquist2008b}
found that a volumetric heating rate of $\epsilon_H\sim B_{avg}/L$, i.e. exponents
$\alpha=1$ and $\beta=1$, provided the best agreement between full active region
simulations made of ensembles of individually modeled loops and X-ray observations.
Note that those studies only considered integers for the exponents. Even more interestingly,
\citet{winebarger2008}, who added constraints from the EUV moss, provided a more precise
estimate of $\epsilon_H\sim B_{avg}^{0.29\pm0.03}/L^{0.95\pm0.01}$ that is in full
agreement with our estimate of $\alpha$. We do not have an explanation for the
discrepancy with $\beta$, but given the distribution of data points in
Figure~\ref{fig:intvsBavg}, with the two outliers at large intensities, it seems
desirable to extend the relationship up to brighter loops to check if the slope
becomes steeper or has multiple components.

There are other studies that are more
difficult to compare with because their prescription of the heating depends on the
footpoint field strength ($B_f$). \citet{schrijver2004}, in their full Sun models
of the corona extending work in \citet{schrijver2002}, find that a heating flux
density $F_H$ ($\rm ergs \,cm^{-2} \,s^{-1}$) that scales as $B_f/L$ provides the
best match to X-ray and EUV observations. As $\epsilon_H\sim F_H/L$, then
$\epsilon_H\sim B_f/L^2$. We have looked at the relationship between $B_{avg}$
and $B_f$ for our dataset and find $B_f\sim B_{avg} L^{0.56\pm0.08}$. Replacing,
that would mean $\epsilon_H\sim B_{avg}/L^{1.4}$, which is not consistent with
our estimates given the uncertainties. \citet{dudik2011} looking at full active
regions and allowing for a spatially variable heating scale-length provide an estimate of
$\epsilon_H\sim B_f^{0.7-0.8}/L^{0.5}$, i.e. $\epsilon_H\sim B_{avg}^{0.7-0.8}/L^{0.1}$
potentially consistent with our results if their uncertainties are of the same order.
While the overall picture does not give us a complete agreement between all the various
estimates, the trend of these results suggests that the volumetric heating rate
scales with $B_{avg}^{\alpha}/L^{\beta}$ with exponents in the range 0.2--1.0.

\section{Conclusion}
We have investigated the magnetic ($B_{avg}$ and $L$) and radiative (\ion{Fe}{18} 93.9\AA\ intensities)
properties of 34 loops from three active regions (NOAA 11158, 11190, 11339), with total unsigned
fluxes spanning $4.2-26\times10^{21}$ Mx, observed with the AIA and HMI instruments on board the {\it SDO} mission.
These loops are formed at temperatures close to the peak of the emission measure in these active regions.
We find that the peak intensity per unit cross-section of observed heating events in the loops
scales proportionally with average magnetic field strength ($\sim B_{avg}^{0.9\pm0.3}$)
and inversely with the loop length ($\sim L^{-0.6\pm0.2}$). We do not find any differences between
loops from different regions. Furthermore, the intensities are uncorrelated with the field strength
at the footpoints.

Our investigation shows that the relationship between intensity, field strength and loop
length is compatible with impulsive heating with a volumetric heating rate
$\epsilon_H\sim B_{avg}^{0.3\pm0.2}/L^{0.2\pm^{0.2}_{0.1}}$. This result, obtained at the scale
of individual loops, is compatible with several previous estimates obtained from full
active region modeling as an ensemble of loops. While pinning down the scaling for
$\epsilon_H$ may be a testable property of heating mechanisms as discussed
by \citet{mandrini2000} \citep[see also][]{lundquist2008b}, the difficulty in finding
ideal systems in the corona makes us think that the actual observables will become
more powerful discriminators.

These results set new observational constraints for the major efforts currently
underway in developing and testing the ability of state-of-the-art numerical experiments to reproduce
observables in the corona. In combination with other constraints such as the relationship
between morphology and magnetic topology, emission measure distributions, short and
long-term variability, loop cross-sectional properties, Doppler-shifts, etc. they are
setting a challenging, but definite description of the plasma properties in the
corona that numerical renderings of the theories need to keep up with.


\acknowledgments We would like to thank the anonymous referee for all the
suggestions that helped improve the paper.
I.U.U. acknowledges support by NASA under Grant No. NNH17AE96I
issued through the Heliophysics Grand Challenge Research Program. H.P.W., N.A.C.
were supported by the Chief of Naval Research and NASA’s Hinode program.
T.W. acknowledges DFG-grant WI 3211/5-1. The authors would like to thank Markus
Aschwanden for the comments and discussions on this work. AIA and HMI data are
courtesy of NASA/{\it SDO} and the AIA and HMI science teams.

\vspace{5mm}
\facilities{{\it SDO} (AIA,HMI)}
\software{CHIANTI \citep{dere1997,delzanna2015}, SolarSoft \citep{freeland2012}}


\bibliography{astroph}
\bibliographystyle{aasjournal}

\end{document}